\documentclass[
showpacs,
aps,prd,
nofootinbib,
superscriptaddress,
tightenlines,
amsmath,
amssymb
]{revtex4}
\usepackage{bm}
\usepackage{color,graphicx}
\usepackage{amsmath}
\usepackage{amssymb}

\begin{document}
\title{Self-Organizing Maps Parametrization of Deep Inelastic Structure Functions with Error Determination.} 

\author{Evan M. Askanazi} 
\email{ema9u@virginia.edu}
\affiliation{Department of Physics, University of Virginia, Charlottesville, VA 22901, USA.}

\author{Katherine A. Holcomb} 
\email{kholcomb@virginia.edu}
\affiliation{University of Virginia Alliance for Computational Science and Engineering, University of Virginia, Charlottesville, VA 22901, USA.}

\author{Simonetta Liuti} 
\email{sl4y@virginia.edu}
\affiliation{Department of Physics, University of Virginia, Charlottesville, VA 22901, USA.}
\affiliation{Laboratori Nazionali di Frascati, INFN, Frascati, Italy}


\begin{abstract}
We present and discuss a new method to extract parton distribution functions from hard scattering processes based on an alternative type of neural network, the Self-Organizing Map.  Quantitative results including a detailed treatment of uncertainties are presented within a Next to Leading Order analysis of inclusive electron proton deep inelastic scattering 
data. 
\end{abstract}
\maketitle

\section{Introduction}
%%%% SOMs
In the past twenty years Artificial Neural Networks (ANNs) have established their role as a remarkable computational tool in high energy physics analyses. 
Important applications have been developed that provide, for instance, classification methods for off-line jet identification and tracking, non-classification-type tools for on-line process control/event trigger and mass reconstruction, and optimization techniques in {\it e.g.}  track finding \cite{Pet}. 
More recently, ANNs have been used to extract Parton Distribution Functions (PDFs) from high energy scattering processes. 
In a series of papers \cite{Ball.2010,DelDebbio.2007,Forte.2002} the authors developed the 
Neural Networks PDFs (NNPDFs), which are PDFs obtained from a neural network based analysis with ``faithfully statistical, systematic and normalization errors" from global fits to data.
ANNs analyses differ from standard global analyses in that they avoid the bias that is associated with the choice of parametric functional forms for the PDFs. 
Each PDF is, in fact, parameterized with a redundant functional form given by a neural network with 37 free parameters represented by the ANN's weights. 
The parametric forms are subsequently
fitted to the experimental data, and their $\chi^2$ minimized, by using a genetic algorithm approach \cite{DelDebbio.2007}. Attention must be paid at this stage to 
the statistical aspects of the approach, and in particular of the synthetic data used in the training.
Several estimators where studied in \cite{Ball.2010,DelDebbio.2007,Forte.2002} 
to assess the quality of the ANN training. These include a ``convergence condition", or ``stopping criterion", which marks the duration of the training phase by the onset of a stage where the neural network begins to ``overlearn", or to reproduce the statistical fluctuations of the data rather than their underlying physical law.
 
In a nutshell, what distinguishes NNPDFs from other methods is, besides an improved statistical analysis, their automated minimization procedure.
The NNPDF approach has the advantage of being a ``black box", which implies that the user's bias in extracting PDFs from experimental data is eliminated.
In other words, no physical assumption goes into determining the shape of the parametrization, rather the physical behavior is inferred directly from the data
in terms of smooth PDF curves. 
This turns out to be disadvantageous, however, in  
kinematical regions where there is no experimental information, or in between the data points if the data are 
sparse. The inherent unbiasedness of this approach also implies that the behavior of NNPDFs in kinematical regions with few or no data 
cannot be sensibly extrapolated from their behavior in regions where data exist. In other words,  since for NNPDFs the effect of modifying individual
NN parameters is unknown -- the weights are in a non tunable, hidden layer --  the result might not be under control
in the extrapolation region.
In summary, ANNs do not work efficiently, or they have a low performance, for extrapolating or predicting the behavior of 
structure functions. 
Yet this aspect of the performance of a fit to high energy physics data is often a desirable one.

In particular, a new generation of experiments have been intensively studied in recent years 
that include high energy, both unpolarized and polarized -- deeply virtual -- semi-inclusive and exclusive scattering processes off hadronic targets ( see Refs.\cite{Diehl_rev,BelRad} for a review). 
Due to their exclusive nature, these measurements are appreciably more challenging to analyze than the inclusive-type ones, since they involve a larger number of observables that in turn depend on several kinematical variables. As a result of the increased dependence on these variables their kinematical coverage has been determined so far spottily, and the observables have been measured with less accuracy.  

With the future goal of  fitting a larger class of experimental data than the unpolarized, inclusive data analyzed with ANN methods so far \cite{Ball.2010,DelDebbio.2007}, 
in this paper we study an alternative approach. We propose Self-Organizing Maps (SOMs) \cite{Kohonen} as an alternative to the ANN-based method used by {\it e.g.} \cite{Ball.2010,DelDebbio.2007}. SOMs are also ANNs where the basic learning algorithm is replaced by an {\em unsupervised learning} approach that makes use of
the underlying {\em self-organizing} properties of the network's input vectors -- in our case the PDFs. Unsupervised learning
works without knowledge about what features characterize the output:
data are organized according to certain attributes without any ``teaching''. 
A most important aspect of the SOM algorithm is in its ability to project high dimensional {\it input} data onto lower dimensional representations while preserving the topological features present in the {\it training} data. 
Because results using unsupervised learning are most often represented as 2D geometrical configurations 
%where a neighborhood of nodes get simultaneously updated while reproducing the clustering of the data's features, 
the new algorithm is defined as a ``map".   

Pioneering work using SOMs for the analysis of high energy physics data was discussed in Ref.\cite{Lonn}.
An initial study of Deep Inelastic Scattering (DIS) structure functions using SOMs was performed in Ref.\cite{Carnahan}, where 
a new approach, SOMPDF, was presented as an alternative to the purely automated fitting procedure of NNPDF. Initial results
were aimed at proving the viability of the method as a fitting procedure, by successfully generating $\chi^2$ values that decreased
with the increasing number of iterations. These studies did not focus, however, on the specific clustering properties of the map.

%The main point is that since for NNPDFs the effect of modifying individual
%NN parameters is unknown, the result might not be under control
%in the extrapolation region, or in between the data points if the data are sparse. 
Here we start from a similar perspective as in the SOMPDF approach of  Ref.\cite{Carnahan},  but we introduce significant restructuring in the original code.  The main purpose of restructuring, described in detail in Section \ref{sec3}, has been to allow us to go beyond ``developing and observing the unconventional usage of the SOM as a part of an optimization algorithm" \cite{Carnahan}, and to actually provide a quantitative analysis of PDFs.

The main modification that we introduce is in the initialization procedure. In our new approach we perform random variations of the PDF {\em parameters} which form the initial set, instead of variations on the {\it values} of PDFs at each value of $x$ and $Q^2$ of the data. Rewriting the initialization stage according to this criterion (Section \ref{sec3}) allows us to obtain smooth or continuous solutions, similar to other global analyses.
In addition, we can now apply a fully quantitative error analysis to our extracted PDFs.
Another important outcome of our new analysis is that the new method offers sufficient flexibility so that it can be applied to different multivariable dependent observables, including  the matrix elements for deeply virtual exclusive and semi-inclusive processes.
Our first quantitative results for the unpolarized case using Next-to-Leading-Order (NLO) perturbative QCD  were presented in Ref.\cite{Perry_dis10,Hol_exc}. Here we present a parametrization of the PDFs at NLO with calculated uncertainties from the SOMPDF method.
  
The paper is organized as follows: in Section \ref{sec2} we review the SOMs, the PDFs and the application of SOMs to PDF fitting (SOMPDFs); in Section \ref{sec3} we present our first set of SOMPDFs (SOMPDF.1) as a quantitative parametrization of PDFs; 
in Section \ref{sec4} we describe in detail our new results. 
%In Section \ref{sec4} we present phenomenological applications. 
Finally, in Section \ref{sec5} we draw our conclusions 
and we discuss the
extension to multi-variable distributions such as Generalized Parton Distributions and Transverse Momentum Distributions. 

%%%%%%%%%%
%%%%%%%%%%
%%%%%%%%%% SECTION II
%%%%%%%%%%
\section{SOMPDF}
\label{sec2}
How do SOMs apply to PDF analyses?  The basic SOM algorithm can be defined as a {\em non linear} extension of Principal Component Analysis (PCA) \cite{PCA}. 
In PCA one applies an orthogonal transformation to convert a set of data that are possibly correlated into sets of values that are linearly uncorrelated, and which constitute the principal components. 
The first principal component exhibits the largest variance, {\it i.e.} it is a straight line that minimizes the sum of the squared distances from the data points (least squares approximation) of the data set by a line. The second principal components is by subtracting from the original data vectors their projections onto the first principal component and by finding a new straight line approximation. The procedure is applied to the following components recursively.
PCA is useful for interpreting the behavior of high dimensional data because, by allowing one to represent the dominant data sets in a linear form, 
and by simultaneously discarding the sub-dominant components, PCA can reduce the number of dimensions of the problem. 
However, PCA cannot account for nonlinear relationships
among data. Furthermore, it has  poor visualization properties in cases where more than two dimensions are important.

The essential feature that sets the SOM algorithm apart from PCA and similar data reduction methods is that the lines resulting from PCA can be effectively replaced by lower dimensional manifolds in the SOM method. Because of their flexibility, these can catch features of the data that the PCA would not.
In addition, SOMs have enhanced visualization features to represent higher dimensional data, while visualization for more than four components becomes an impossible task for PCA \cite{Haykin}. 

Finally, from the theoretical point of view, SOMs are particularly relevant algorithms in systems theory, as they model the emergence of a collective ordering in a composite system through the competition of its constituents. We can foresee a number of future applications to complex nuclear and high energy data using this aspect of the SOM method \cite{Ireland}.

Below we summarize the PDFs fitting procedures, the SOMs algorithm, and we subsequently describe how we match the two. 

\subsection{PDFs}
PDFs, $f(x,Q^2)$, with $f=q, \bar{q}, g$, describe the structure functions of deep inelastic processes in QCD. In the $\overline{\rm MS}$ scheme one writes,
\begin{eqnarray}
F_2(x,Q^2) & = & x \sum_{q,\bar{q}} e_q^2 \int_x^1 \frac{dy}{y} C_q(y,\alpha_S) \, q\left(\frac{x}{y},Q^2\right)   
+  x  \int_x^1 \frac{dy}{y} C_g(y,\alpha_S) \, g\left(\frac{x}{y},Q^2\right),
\end{eqnarray}
where $C_{q,g}$ are the coefficient functions, $\alpha_S(Q^2)$ is the running strong coupling. PDFs are obtained in most global fits at least at next-to-leading order (NLO) in Perturbative QCD (PQCD);  next-to-next-to-leading order (NNLO) evaluations are performed in most cases by using data sets from processes for which theoretical calculations to that order exist \cite{bench}.
%\begin{eqnarray}
%\frac{\partial}{\partial \ln Q^2} f_i(x,Q^2) = \frac{\alpha_S}{2 \pi} \int _x^1 \frac{dy}{y} P_{qq}\left(\frac{x}{y}, \alpha_S\right) f_i(y,Q^2) \nonumber \\
%\end{eqnarray}
%
In particular, PDFs are extracted up to NNLO from deep
inelastic lepton-nucleon scattering, and  up to NLO from  hard scattering processes in an ever growing range of $x$ and $Q^2$ -- from the large $x$ multiGeV region
fixed target measurements at Jefferson Lab \cite{Acc1,Acc2,CouLiu,BiaFanLiu}, to the range of LHC precision measurements of $W^\pm$, $t\bar{t}$.
The continuously increasing experimental data set which needs to be included in the analyses while simultaneously attacking the various theoretical open questions in PQCD make global fits an exacting enterprise.
Currently, several groups have determined the parameterizations for the unpolarized PDFs.
MSTW \cite{MSTW}, CT \cite{CT10}, HERAPDF  \cite{HERAPDF}, ABM \cite{ABKM,ABM} and
JR \cite{JR1,JR2} use a parametric form for the PDFs with $4-5$ free parameters per parton distribution type, 
at an input scale, $Q_o^2$,  which varies for the different fitting forms. 
NNPDF \cite{Ball.2010,DelDebbio.2007,Forte.2002} 
use neural networks to determine the initial
input distributions in an unbiased way.  
%All parametrizations use PQCD evolution equations to Leading Order (LO), Next to Leading Order (NLO), and Next to Next to Leading Order (NNLO) in $\alpha_S$.
A summary of all current PDFs parametrizations and their uncertainties is given in \cite{bench}.

%%%%%%%%%%%%%%%%
\subsection{SOM algorithm}
%% Nodes
The SOM is formed by a two dimensional $n\times n$ grid of neurons, or nodes.\footnote{In our case we choose a square map, other topologies are possible \cite{Kohonen}.}
% Input Vector
The nodes are presented with a stimulus, parameterized in a vector of dimension $n$; this is called the input vector and it describes the set of data to be processed.   
%% Weight Vector
Each element of the vector is presented to all nodes on the map with a synapse or weight, $V$.  Each node corresponds to the weight vector $V$ containing $n$ weights (same dimension as the input vector).

The SOM algorithm consists of three stages: {\it i)} Initialization; {\it ii)} Training; {\it iii)} Mapping. 

\subsubsection{Initialization}
\label{sec:init}
The SOM learning process is defined as {\em unsupervised} and {\em competitive}.  
During the initialization procedure weight vectors of dimension $n$ are associated with each cell $i$:
\[ V_i = [v_i^{(1)}, ..., v_i^{(n)} ]  \] 
$V_i$ are given spatial coordinates, {\it i.e.} one defines the geometry/topology of a 2D map
that gets populated randomly with $V_i$.  Typically each of these vectors consist of a randomized value of the type of data that is to be represented.  We will generally
refer to these initial vectors as {\it map vectors}.

\subsubsection{Training}
\label{sec:train}
Next, an input vector is presented to the grid. The node whose map vector is most similar to the input vector is defined as the {\it best matching unit} or {\it BMU}. The weights, $V$, of the BMU and of the surrounding nodes form a neighborhood, $N$, of some specified radius $r$.  
The weights of this set of map vectors are then modified so that the map vectors move closer, according to some chosen metric, to the input vector.  After all nodes have been adjusted the neighborhood radius is reduced according to some 
criterion and the training procedure is repeated.  
%Specifically, the algorithm reads
%\begin{subequations}
%\begin{eqnarray}
%w_i(t+1) & = & w(t) + h(r,t) (p_i-w_i(t)) \; \; \;  {\rm for} \;  i = N   \\
%w_i(t+1) & = & w(t) \; \; \;  {\rm for}  \; i = N
%\end{eqnarray} 
%\end{subequations}
%with $h=\alpha v$, $\alpha$ being a learning coefficient and $v$ the neighborhood function. 
%KAH Need to say somewhere earlier that this is iterative.  Why is previous stuff commented out?

For the training, a set of input data
\[ \xi = [\xi_i^{(1)}, ..., \xi_i^{(n)} ] , \] 
(isomorphic to $V_i$) is then presented  to $V_i$, or compared  via a {\it similarity metric} that we choose to be,
\begin{equation}
\label{metric}
 L_2(x,y) = \sum_{i=1,2} \sqrt{  x^2_i - y^2_i  }.  
\end{equation}
This is the ordinary Euclidean norm for vectors $x$ and $y$. 

%SOMs are based on  unsupervised and {\it competitive} learning.
The unsupervised  part of SOMs training takes place as the cells that are closest to the BMU activate each other in order to ``learn" from $\xi$.  
Practically, they adjust their
values according to the following algorithm,
\begin{equation}
V_i(n+1) = V_i(n) +  h_{ci}(n) [\xi(n) - V_i(n)],
\label{learn_1}
\end{equation}
where $n$ is the iteration number, and $h_{ci(n)}$ is the {\it neighborhood function} defining a radius on the map which decreases with both $n$, and
the distance between the BMU and node $i$. In our case we use square maps of size $L_{MAP}$, and 
\begin{equation}
h_{ci}(n) = 1.5 \left(\frac{n_{train} -n}{n_{train}} \right) L_{MAP}
\label{learn_2}
\end{equation}
where $n_{train} $ is the number of iterations.
At the end of a properly trained SOM, cells that are topologically close to 
each other will contain data which are similar to each other.
In the final phase the actual data are distributed on the map and 
clusters emerge.  Note that the specific location of the clusters on
the map is not relevant and will not necessarily be the same from one run to another;
only the clustering properties are important.
%The values of $v$ and $\alpha$ are then decreased until the learning process is stopped when $\alpha=0$. This algorithm based on competitive learning, is also called vector quantization.

Once the learning process is complete, each new set of data will be associated with the location of
its BMU.   

\subsubsection{Mapping}
\label{sec:map}
SOMs are built as two dimensional arrays  whose cells get sensitized/tuned to a specific set of input signals according to a given order.  
Since each map vector now
represent a class of similar objects, the SOM
is an ideal tool to
visualize high-dimensional data, by projecting it onto a low-dimensional map
clustered according to some desired similarity feature. 

\subsection{Representing PDFs as SOMs}
\label{sec2C}
In our analysis the vectors are sets of candidate PDFs, $V_i\{{\bf x}_n\}  \rightarrow f_i(\{{\bf x}_n\},Q_o^2)$, $f=q, \bar{q}, g$, which are randomly generated to form an initial envelope. 
$\{{\bf x}_n\}$ is a vector of $n$ Bjorken $x$ values; each PDF is evaluated at the corresponding $x$ values, at the initial scale, $Q_o^2$.  

We select PDFs from the envelope to: {\it i)}  generate training data, the {\it code vectors}; {\it ii)}  place vectors on the map, the {\it map vectors}.
An iteration is defined as the process where the entire set of code vectors, or input PDFs, is presented to the map vectors, the most closely matching the input PDFs being declared the ``winning" PDFs. The map PDFs are organized around the winning one according to the algorithm described in Section \ref{sec:train}. 
Various map parameters need to be fixed at this point including the size and shape of the map, the number of iterations, the number of PDFs used in each training cycle, the initial learning rate, etc.. The map parameters values are presented and discussed in detail in Section \ref{sec3}.

After the map is trained 
%\subsection{Genetic Algorithm}
%\label{sec2D}
we use a  Genetic Algorithm (GA) whereupon the new map PDFs, or the input PDFs, are analyzed relative to known experimental data for the observable -- $F_2$, in this case -- values. 
At this point we perform PQCD evolution to the $Q^2$ values of the data by evaluating the PDFs' Mellin moments (details on this step are given in Section \ref{sec3}).

It is important to underline the difference between the {\it experimental data} and the {\it training data}, as this defines unsupervised learning. 
The training data constitute a bundle/envelope of possible PDF curves which encompasses the measured data. Elements of the training data set are identified with the map vectors which are built in the initialization part. 
A  $\chi^{2}$ is evaluated for each (evolved) map cell; PDFs with the lowest $\chi^{2}$ values are used as seeds for the next set of input PDFs.  This process is repeated for $N$ iterations; over the course of these iterations the $\chi^{2}$ values eventually asymptotically approach a given value, which is referred to as the saturation value. Reaching the saturation values defines the ``stopping" criterion for the fitting procedure in our case.

%%%%%%% SECTION III
%%%%%%%
\section{SOMPDF as a quantitative parametrization of DIS data}
\label{sec3}
In this Section we describe the details of our fitting procedure. Several aspects of our approach were defined in an initial exploratory study \cite{Carnahan}.
In order to perform a fully quantitative fit  various changes were taken into account which we describe in detail below. 

%%%
\subsection{New Initialization Method}
\label{sec3A}
We start by describing the construction of the initial envelope for the SOM training. When we subsequently apply the GA, we construct new envelopes which contain sets of PDFs that are generated from each previous iteration, and that are selected based on their $\chi^{2}$ values so that, after a number of  iterations we minimize it.  

The challenge one meets in forming an initial envelope  is that on the one hand it must be constructed randomly in order to meet the criterion of unbiasedness, and on the other hand it must be adjusted sufficiently enough to somewhat loosely follow the experimental data.  

Our envelope is formed by taking three different parametric sets of  PDFs,  given in \cite{JR1,JR2,MSTW, ABKM}, at an initial $Q^2= Q_o^2=0.58$ GeV$^2$. 
%We then introduce $Q^2$ dependent parameters by using the procedure from GRV \cite{GRV}.  $Q^2$ dependent parameters were introduced for all three parametric forms we used. 
In order to illustrate how the parameters were varied in order to form an envelope we take, for example, the NLO JR parametric form \cite{JR1} at $Q_o^2 = 0.34$ GeV$^2$,
%
%\begin{subequations}
\begin{eqnarray}
\label{qval}
x f_i & = & A_i \, x^{B^i_1} (1-x)^{B^i_2} F_i(x)
\end{eqnarray} 
%\end{subequations}
where $i=u_v, d_v, \bar{u}, \bar{d}, s, c, g$, and $F_i(x)=(1 + C^i_1 \sqrt{x} + C_2^i \, x) $ guarantees more flexibility in the functional form.
The parametrization of  the initial PDFs in \cite{MSTW, ABKM} differ in the functional forms for $F_i(x)$, which are given by {\it e.g.} combinations of polynomials and exponentials.

The only constraints that were imposed on the PDFs at this stage are from the baryon number and momentum sum rules,
\begin{eqnarray}
\label{baryon}
&& \int_0^1 dx \, u_v  =  2, \;\;\;\;\;\; \int_0^1 dx \, d_v = 1 \\
\label{momentum}
&& \int_0^1 dx x \, \left[(u_v + 2 \bar{u}) + \left( d_v + 2 \bar{d} \right) + 2 \bar{s} + 2 c + g \right ] = 1
\end{eqnarray}
The observables of interest in this paper are the DIS proton and neutron electromagnetic structure functions, $F_2^{p,n}$, 
\begin{eqnarray}
F_2^p & = & x \left[ \frac{4}{9} \left( u_v + 2 \bar{u} \right) + \frac{1}{9} \left( d_v + 2 \bar{d} \right) 
          +  \frac{2}{9} s +  \frac{8}{9} c   \right] \nonumber \\ \\
          F_2^n & = & x \left[ \frac{4}{9} \left( d_v + 2 \bar{d} \right) + \frac{1}{9} \left( u_v + 2 \bar{u} \right) 
          +  \frac{2}{9} s +  \frac{8}{9} c   \right], \nonumber \\
\end{eqnarray}
where the intrinsic charm component starts at $Q^2 \geq m_c^2$. The neutron structure function is extracted from deuteron experimental data using  $F_2^d=F_2^p+F_2^n$. The data sets are described in Section \ref{sec:data}.

The parameters initial values were set to be consistent with the ones obtained in Refs.\cite{JR1,MSTW, ABKM}. 
Notice that details of the fit in {\it e.g.}  Ref.\cite{JR1}, or even the fact that the PDFs given above can properly fit existing data, are not important for constructing the envelope. It is essential, however,  to have parameterizations that provide functional forms sufficiently close to the data so that, by properly varying some of their parameters, one can construct a bundle of curves whose envelope encompasses all of the available data. 
This step of our analysis can be challenging in that by using some of the baseline parametrization formulas it might be harder to bracket newer data.

%%
%We determined that when we started from the baseline GRV formulas it was nearly impossible to bracket newer data.  Since there is no particular reason to start from
%an exact GRV expression we made several adjustments.  
%WHAT WAS EXACTLY DONE AND WHIC PARAMETERS WERE WIGGLED. Example of envelope.
In order to estimate the parameters' variations that were necessary to form a statistically meaningful bundle of curves (envelope), we took the initial values of the parameters, using a $4 \times 4$ map,  and we varied them according to the following rules:
 
{\it i)} Firstly, we found that several parameters had very little impact on the range of the overall result when they were varied within 
reasonable ranges, so we left them fixed. 
Certain parameters were not
varied since they would too easily drive the computed PDF to overflow without a compensating improvement in the envelope of the resulting PDFs.  

{\it ii)} All of the remaining  parameters were varied so as to accomplish the best 
bracketing of our entire set of experimental data.  
%These adjusted parameters became the new baseline around which we varied.  
In particular, the parameters of interest were
%%%% EA
all the exponential parameters. 
%namely  $B^q_1$, $B^q_2$,  $\alpha_q$, $D_q$.
In order to compute a ``vector'' for the SOM, we multiplied each parameter by a random variable from a normal distribution with a mean of $1$ 
and a specified standard deviation, 
\begin{equation}
\label{random1}
P^{new} = P \pm \Delta P, 
\end{equation}
where $P$ is any of the parameters listed above, and $\Delta P$ is obtained from the distribution with $\sigma = 0.1$.
This procedure was repeated using the parametric forms from Refs.\cite{ABKM,MSTW}.   We then constructed a weighted sum of the three ``wiggled''
functions, to generate each PDF/``vector" in the envelope, 
\begin{equation}
\label{random2}
f_i^{env} = C_1 f^1_i + C_2 f^2_i + C_3 f_3^i,
\end{equation}
where the index, $i$, indicates the partonic component; each coefficient, $C_j$, $j=1, 2, 3$, is a uniform random number; $f^1_j$ are the PDFs from Ref.\cite{JR1} ($j=1$), Ref.\cite{ABKM} ($j=2$), and Ref.\cite{MSTW} ($j=3$), respectively, which get randomized using Eq.(\ref{random1}).  
Each randomized PDF was normalized so as to obey the sum rules in Eqs.(\ref{baryon},\ref{momentum}).  

In order to obtain all parametrizations at a common initial $Q_o^2$, we explicitly introduced a $Q^2$ dependence of  the parameters  on,
\begin{equation}
s = \log \frac{\log Q^2/\Lambda^2}{\log Q_o^2/\Lambda^2} 
\end{equation}
as in early PDF parametrizations (see {\it e.g.} \cite{GRV} and references therein). This gives us for instance, $A_i \equiv A_i(s), B_i\equiv B_i(s), ...$, in Eq.(\ref{qval}). Again,
it should be kept in mind that in order to construct the envelope, precision is not a requirement, while it suffices that the input function need just to encompass the data. We chose
$Q_o^2 = 0.58$ GeV$^2$. Variations of this value do not affect sensibly our results. 
% as it can be seen, for instance, in Fig.\ref{fig:envelope} where we give an example of one of our envelopes.
%%%% 
%%%% FIGURE 1 ENVELOPE 
\begin{figure}
%\includegraphics[width=7.cm]{nnpdf_example.pdf}
%\hspace{-4.5cm}
%\includegraphics[width=9.cm]{pdfenvelope.pdf}
\includegraphics[width=8.5cm]{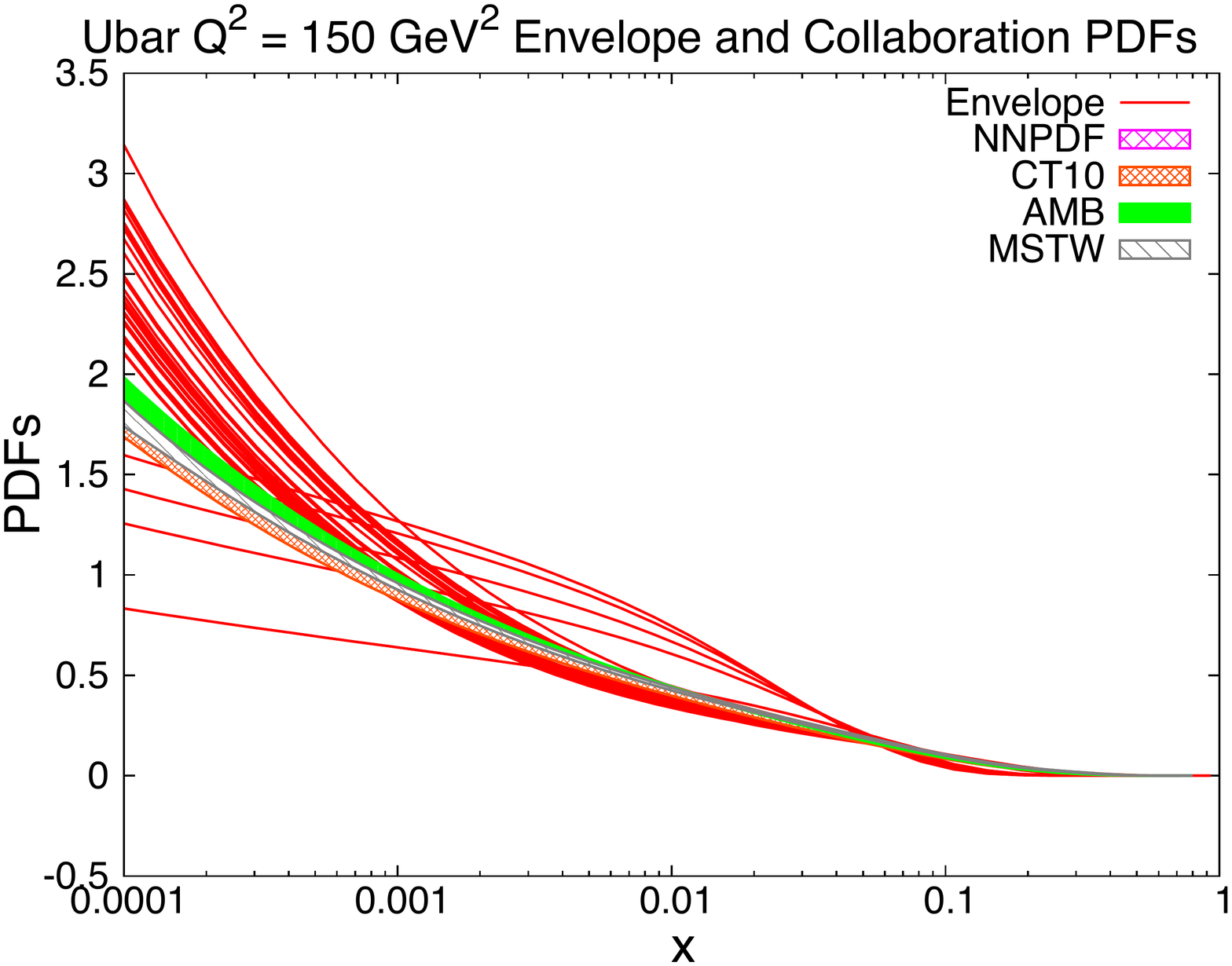}
\includegraphics[width=8.5cm]{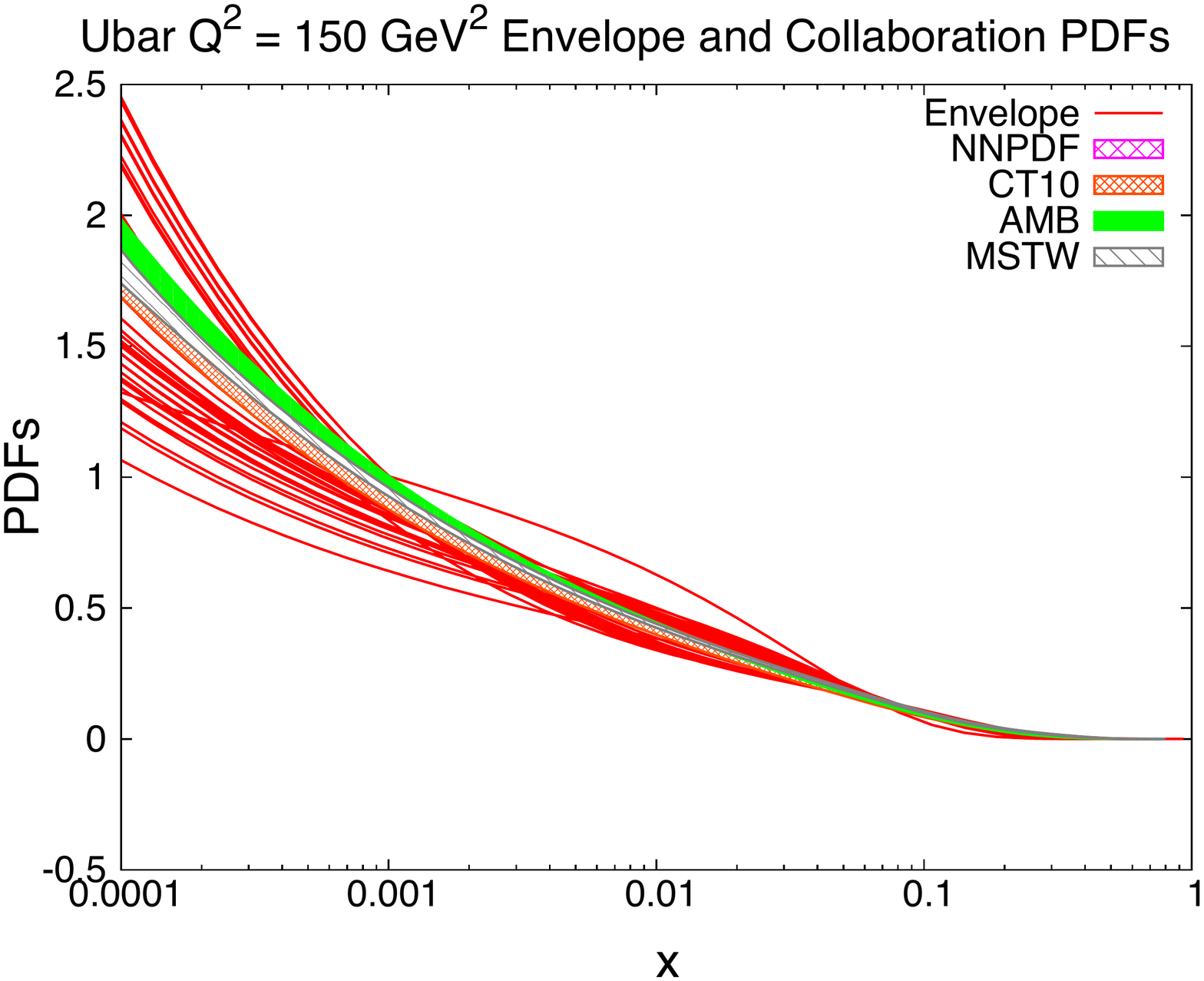}
\caption{(Color online) Example of an envelope characterizing the SOMPDF initialization stage for the $\bar{u}$ PDF. 
On the LHS we show the spread of curves for the initial GA step; on the RHS we show the same spread but at the final GA step (in this case for $N_{MAX}=200$). Similar results are obtained for all the other PDFs. 
Results from different PDFs fits \cite{Ball.2010,ABM,CT10,MSTW}, are shown for comparison.}
\label{fig:envelope}
\end{figure}
%%%%%%%%%%%%%%%%%%%%%%%%%%%
The resulting envelope of PDFs that effectively wrapped around the experimental data from above and below  
minimizes the  bias in determining the lowest $\chi^2$ values for use in the GA.
The vectors/each generated PDF now contain a mixture of up to three different $x$ and $Q^2$ dependent parametric forms whose parameters were varied in order to assure both randomness and the observance of physical constraints.  A set of these makes up a vector for a particular cell; the number generated for each cell is a parameter of the code.  Once the envelope is formed, we proceed with the training and GA.
Examples of envelopes for the $\bar{u}$ distribution are given in Fig.\ref{fig:envelope}.

%%%% ENVELOPE
%\noindent \underline{\it Envelope}:
%Figure \ref{fig:envelope} describes the . Shown in the figure are also the ``confidence levels" from our error analysis (Section ...). 

%In addition, if the non exponential parameters were scaled too much in the attempt to create a converging envelope, there was a risk of the pdfs being forcefully reverted to their default values, making it impossible to create the envelope. 

\subsection{PQCD Evolution: Moments}
Because of the increased flexibility allowed by the new initialization procedure,  we were able to introduce also a new, more flexible criterion 
to take into account $Q^2$ evolution. Although in this paper we limit our evaluations to PQCD at NLO, our new method allows us to naturally take into  
account other sources of $Q^2$ dependence from {\it e.g.} NNLO, target mass corrections and higher twists that affect for instance the large $x$ behavior of the structure functions. 

Perturbative evolution was taken into account in Mellin space, the Mellin moments being defined as \cite{Roberts},
\begin{equation}
M_n(Q^2) = \int_0^1 dx x^{n-1} f_i(x,Q^2).  
\end{equation}  
%for $f=q,\bar{q,g}$.
For the non singlet combinations, the PQCD dependence of Mellin Moments is given by  
\begin{eqnarray}
M^{NS}_{n}(Q^{2}) = M^{NS}_{n}(Q^{2}_o) \left(\frac{\alpha_{s}(Q^{2})}{\alpha_{s}(Q^{2}_o)}\right)^{d^{NS}_{n}}  \left[1+ C_n^{\overline{MS}}\left(\frac{\alpha_{s}(Q^{2})}{4 \pi} -\frac{\alpha_{s}(Q_o^{2})}{4 \pi}\right)\right] 
\end{eqnarray}
where ${d^{NS}_{n}}= \gamma_{qq}^n/2\beta_0$, and $C_n^{\overline{MS}}$ is given in \cite{Roberts}. 
%refers to the non singlet constants of asymptotic freedom given in Roberts \cite{Roberts}.

The singlet structure function form is much more involved due to the coupling to the gluons, and we do not report it here (see \cite{JR1} for the full expressions). 
%For $n=2$ one has, 
%One has,
%\begin{eqnarray}
%\left[ M^{S}_{n}(Q^2) - \frac{3N_f}{16 + 3N_f}   \right] = \left[M^{S}_{n}(Q^2_o) - \frac{3N_f}{16 + 3N_f } \right] \left( \frac{\alpha_s(Q^2)}{\alpha_s(Q^2_0)}\right)^{d^+_2}   
 %\left[1+ C_n^{\overline{MS}}\left(\frac{\alpha_{s}(Q^{2})}{4 \pi} -\frac{\alpha_{s}(Q_o^{2})}{4 \pi}\right)\right]  . 
%\end{eqnarray}
 
The value of $\alpha_S(Q^2)$ is obtained at NLO  by solving,
\begin{eqnarray}
%   D & = &   \frac{1}{9.0 (\log \frac{Q^2}{0.30} ) } -  0.1 (\log(\log(\frac{Q^2}{0.30} )))/(((\log \frac{Q^2}{0.30} ))^{2})) 
\frac{d (\alpha_S/4\pi)}{d \ln Q^2} = - \beta_0\left(  \frac{\alpha_S}{4\pi} \right)^2 - \beta_1 \left(  \frac{\alpha_S}{4\pi} \right)^3
\end{eqnarray}
with $\beta_0 = 11- 2 N_f/3$, $\beta_1 = 102- 38 N_f/3$. We use the procedure of Ref.\cite{Mar} to evolve $\alpha_S$ over the heavy quarks mass thresholds. 
%%%%
%%%% alpha_S
%%%%
The value of $\alpha_S(M_Z)$ is allowed to vary in the range  $\alpha_S(M_Z)= 0.1135-0.1195$ consistently with other PDF extractions. 
The correlation between $\alpha_s$ and the PDFs uncertainty \cite{CT_alpha} is therefore implicitly taken into account in our approach. A detailed study will be presented in \cite{Aska_prep}.  

%%%%
All envelope PDFs are normalized so that they integrate to $M_n(Q^2)$, for $n=1$ (valence), and $n=2$ (all components). 
%The sum of $n=2$ moments gives the momentum sum rule, Eq.(\ref{momentum}).  
The values of the moments are obtained from the same set of PDFs  in the initialization step \cite{MSTW,JR1,ABKM},  
 varying randomly them. 
In Table \ref{table:moments} we show the moments values for  \cite{MSTW,JR1,ABKM}, along with results from \cite{Ball.2010}, and our final results (two variants of the fit, corresponding to two different sizes of the SOM, are explained in Section \ref{sec:min}).

%%%%%% TABLE 1
%%%%%%
\begin{table}
%\center
\begin{tabular}{|c|c|c|c|}
\hline 
%
%%%%% begin 
  Collaboration    & $M_2^{val}$  & $M_2^{sea}$ & $M_2^{G}$ \\ 
\hline
\multicolumn{4}{|c|}{$Q^2= 2.5$ GeV$^2$} \\
\hline
ABM    \cite{ABM}      & 0.4644 &  0.0849 & 0.4226   \\
CT10  \cite{CT10}       &  0.4482  &   0.0873 & 0.4263 \\
MSTW   \cite{MSTW}     &      0.4416 &  0.0900 &   0.4322 \\
NNPDF  \cite{Ball.2010}   & 0.4601&  0.0790 & 0.4378    \\
SOMPDF $6\times 6$ & 0.4903  & 0.0867 &  0.4406 \\
\hline
%%%%%%
\hline 
\multicolumn{4}{|c|}{$Q^2= 150$ GeV$^2$} \\
\hline
%
%%%%% begin 
%  Collaboration    & $M_2^{val}$ (isoscalar) & $M_2^{sea}$ & $M_2^{G}$ \\ 
%\hline
ABM          &  0.3407     &  0.0995 &  0.5098 \\
CT10         &  0.3276     &  0.1010  & 0.4793 \\
MSTW       &   0.3177    &  0.1050   & 0.4844 \\
NNPDF     & 0.3340      &  0.0965 &   0.4861 \\
SOMPDF $6 \times 6$ & 0.3425  & 0.0934 & 0.4849 \\
\hline
\end{tabular}
\caption{%
\label{table:moments} 
Central values of the Mellin moments calculated at $Q^2=2.5$ GeV$^2$ (top) and $Q^2= 150$ GeV$^2$ using a $6\times 6$ map for the SOMPDF fit, compared to other current parametrizations \cite{ABM,Ball.2010,CT10,MSTW} .}
\end{table}
%%%%
%Q^2 = 150 uv + dv
%ALBMV 0.3046
%CTEQ10  0.2899
%NNPDF 0.2955
%MRST  0.2888
%6 x 6 0.2857
% 1 X 1 36 0.2845
%
%Q^2 = 2.49  uv + dv
%ALBMV 0.3904
%CTEQ10  0.3785
%NNPDF 0.3868
%MRST  0.3860
%6 x 6 0.3290
%1 X 1 36 0.3245
The spread in the moments thus obtained gives a theoretical/systematic error in our analysis that adds to the non linear correlations provided by the SOM. 

\subsection{Map Features}
Our procedure allows for a number of map parameters  to be adjusted at each run. We list below the values that represent the best choices in terms
of speed of convergence and flexibility of results, and that were therefore used in our final runs: 
\begin{itemize}
\item size of the SOM which we take as an $n \times n$ map; 
\item number of PDF types to be used for mixing,  $n_{PDF}= 1-3$;
\item number of  PDFs per cell, $n_{cell} =2$;
\item number of PDFs to be generated for each cycle during training,  $n_{gen}=4$;
%  pdfs_gen_per_cell =4
\item number of new PDFs to be generated each cycle, $n_{NEW} =10$
% if different from pdfs_gen_per_cell*grid_dim1*grid_dim2When running in parallel, use pdfs_gen_per_cell
%!  num_gen_pdfs =  10
\item number of steps to be used in training each SOM, $n_{step} =5$;
%  somap_steps = 5
%
%! Whether to retrain at each iteration retrain = .true.
%
\item type of norm  ({\it e.g.} $L_1$, or $L_2$) to use for calculating distances between map and code PDFs, $L_2$, Eq.(\ref{metric});
\item initial learning rate, $L_R^0=1$; 
%  ilr=1.0
%! Cutoff for chi-squared values of PDFs that are to be saved from
%! one iteration to the next.
%  starting_metric_cutoff = 102. ending_metric_cutoff = 2.
%
%! Depending on the generator, the stddev parameter will be the actual
%! standard deviation or will be the fraction by which to multiply parameter
%! Number of copies of the original generator to be used among the
%! generators at each future generation.
%! Not used yet
%  copies_orig_gen = 2
%

%\item whether to include map vectors for determining the best PDFs; 

%! Include map vectors in determining best PDFs  
%  use_maps=.false.
%
\item maximum number of iterations regardless of the fitting method, $N_{MAX}=200$;
\item slope parameter based on the number of previous $\chi^2$ values to look at when checking whether the $\chi^2$ curve had flattened out yet, $s_{flat}= 2 \times 10^{-3}$.  
\end{itemize}
%{\bf Check on ovefitting} if we use Heli's algorithm for checking for overfitting.

In addition, we define a tolerance factor to detect over-fitting.  Similarly to the procedure devised in Ref.\cite{Carnahan}, the 
over-fit detection works by taking alternating points from the given 
PDF,  and comparing its  error to the 
error from the best  PDF.  We conclude that the curve
is being over-fit when the error increases from the best PDF to the
``alternating points" PDF by at least $5\%$.
%%%%%% TABLE 2
%%%%%%
\begin{table}
%\center
\begin{tabular}{|c|c|}
\hline
\hline
%
%%%%% begin 
  Size    & Minimum $\chi^2$ \\ 
\hline
\hline
    $4 \times 4 $                 &  0.903 \\
    $5 \times 5 $                 &  0.887  \\
    $6 \times 6 $                 &  0.852 \\
    $7 \times 7 $                 &  0.832  \\
\hline
%%%%%%
\end{tabular}
\caption{%
\label{table:som} 
Minimum $\chi^2$ values obtained for each size of the SOM.}
\end{table}
%%%%

All of these parameters were studied in different ``experiments". 
 We ran a large number of maps of size $4 \times 4$ with $200$ iterations since we could run each one in a reasonable time; these experiments were used to determine empirically the range of parameters that would be appropriate.  Typically each cell has $2$ GPDs that are part of the SOM and $4$ GPDs that are generated in each of the cycles that form the iterations of the GA.  Parameters of the PDF are varied by multiplication with a normally-distributed random number with mean of $1$ and standard deviation of $1/10$ the magnitude of the original parameter value.  We tried experiments in which a larger standard deviation of $4/10$ and $6/10$ was applied for greater variation but this has not resulted in the $\chi^{2}$ values dropping below $2.5$.
 %, which is too large as ideally we desire a $\chi^{2}$ value between $1.0$ and $1.5$.  
 A smaller $2\times2$ map was run for $400$ iterations but increasing the range of the variations did no significantly lower the lowest possible limit of the $\chi^{2}$ values for the SOMPDF generated PDFs.  Larger maps must be run for a better result, but any map larger than a $4\times4$ is highly time consuming even though our code is parallelized using MPI.  
The $\chi^2$ values per map size are listed in Table \ref{table:som}. An $8\times 8$ map did not improve the $\chi^2$ further. We deduce that ideal sizes of the map are either $7 \times 7$, or $6\times 6$, a result consistent also with the analysis in Ref.\cite{Lonn}. 
%Exp Data Set       # of Points
%BCDMS              178  
%SLAC               194
%ZEUS               242
%H1                 130
%EMC                59
%E665               46
%F2P NMC            136
%F2D NMC            152
%%%%%%%%%%%%%%%%%%%%%%%

%%%%
%%%% DATA
%%%%
\subsection{Experimental data}
\label{sec:data} 
\begin{figure}
\includegraphics[width=9cm]{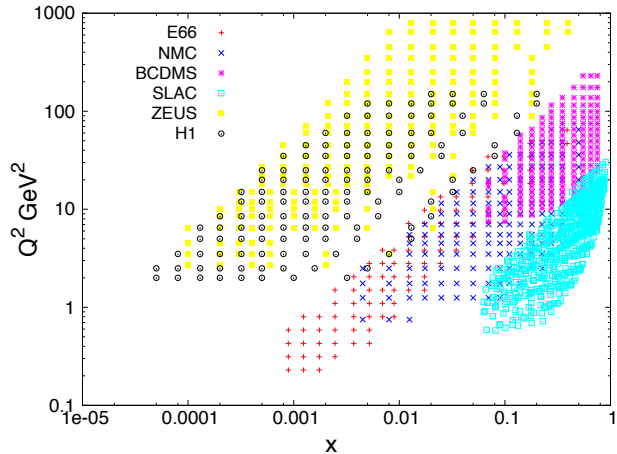}
\caption{(Color online) Kinematical range of the experimental data used in our analysis \cite{SLAC,BCDMS,NMC,E665,H1,ZEUS}.}
\label{fig:kinematics}
\end{figure}

The most recent set of PDF parameterizations has been determined by several collaborations \cite{bench}. Among these Refs.\cite{CT10,MSTW,Ball.2010} use DIS data along with a variety of collider data, while \cite{ABKM,JR1,JR2} focus on  the most precise, or ``highest quality"  DIS data sets only. Similarly, in this first quantitative analysis, we use $ep$ and $ed$ DIS data only, namely the sets from SLAC \cite{SLAC}, BCDMS \cite{BCDMS}, NMC \cite{NMC}, Fermilab E665 \cite{E665}, H1 \cite{H1} and ZEUS \cite{ZEUS}. Our aim is in fact to test the working of our method in a clean way, thus avoiding the complications that necessarily arise due to the treatment of  different types of processes and observables. 
% In Table \ref{table:DIS} we list the data sets implemented in our analysis along with their kinematical range. The data's kinematical range is also shown in Figure \ref{fig:kinematics}.
In Figure \ref{fig:kinematics} we show the kinematical range of the data sets implemented in our analysis.
%
%%%%

%\begin{table}[h]
%\begin{tabular}{|c|c|c|c|c|c|}
%\hline
%\hline
%
%%%%% begin 
%Experiment             &  Measurement                &  Usable Points                & $x$ range       &  $Q^2$ range (GeV$^2$)       \\ 
%\hline
%\hline
%$BCDMS $                 & $F^{p}_{2}$              &     178             &    $7 \times 10^{-2} - 0.75$       & $ 7.5 - 230  $              \\
%$H1$           &         $F^{p}_{2} $                        &  130             &    $1 \times 10^{-4} - 0.85   $    &   $ 2 - 1.5 \times 10^{2}  $             \\
%$ZEUS$     &           $ F^{p}_{2} $                 &            242        &    $ 1 \times 10^{-4} - 0.85   $   &  $ 2.15 - 3 \times 10^{4}$              \\
%$SLAC$                    & $F^{p}_{2} $       &             93               &    $7 \times 10^{-2} - 0.85   $    & $ 0.59 - 31  $             \\
%$NMC$                    &  $F^{p}_{2}, F_2^d$         &    83          &    $8 \times 10^{-3} - 0.5   $      & $ 0.8 - 65 $         \\
%$E665$  &  $F^{p}_{2}, F_2^d$      &    &                                 &                                                       & \\ 
%
%\hline
%%%%%%
%
%\end{tabular}
%
%\caption{%
%\label{table:DIS} 
%Kinematical range of the experimental data used in our analysis.}
%\end{table}

%%%% FIGURE data
%%%%%
%%%%%
The $\chi^2$ is evaluated according to \cite{Pumplin1},
\begin{equation}
\label{chi2}
\chi^2 = \sum_{i_{exp}} \chi^2_{i_{exp}} = \sum_{i_{exp}} \left[ \sum_{j_{data}} \left(\frac{{\cal N}_{i_{exp}} D_{j_{data}}^{i_{exp}} - T_{j_{data}}^{i_{exp}}}{\sigma_{j_{data}}^{i_{exp}} } \right)^2%
+ \left(\frac{1- {\cal N}_{i_{exp}}}{\sigma^{i_{exp}}}  \right) \right]
\end{equation}
where we take into account the correlated error,  $\sigma^{i_{exp}}$, in the normalization of the different data sets, ${\cal N}_{i_{exp}}$ (${\cal N}_{i_{exp}}=1$, for no offset of the normalization). In Eq.(\ref{chi2}), $D_{j_{data}}^{i_{exp}}$, and  $T_{j_{data}}^{i_{exp}}$, refer to the data points (D) and theoretical estimate (T) at each given ($x$,$Q^2$); $\sigma_{j_{data}}^{i_{exp}}$ is the statistical uncertainty. The values of ${\cal N}_{i_{exp}}$ are provided along with the data, for all data sets used in this paper.

\subsection{Error Analysis}
\label{sec:err}
%The fundamental description of the Lagrange method is as follows:
For our error analysis we used the Lagrange multipliers method.
This method evaluates the variation of the $\chi^2$  along a specific direction defined by the maximum variation of a given physical variable. 
In our case the physical variables are the proton (deuteron) structure functions $F_2^{p(d)}$.
However, at variance with previous analyses that used this method \cite{Pumplin1,Pumplin2}, we do not have at our disposal  sets of individual parameters for each given PDF, that can be varied. 
In order to overcome this problem we devised a strategy 
that we describe below, which uses SOMPDFs on appropriately rescaled data.

We follow the application of the Lagrange Multiplier method to PDFs global analyses
outlined in Ref.\cite{Pumplin1} where one takes effective new
$\chi^2$ values determined by, 
\begin{equation}
\label{chi_lambda}
\chi^2(\lambda) = \chi^2_o + \lambda F_2^{p(d)} 
\end{equation}
where
$\lambda$ is a series of Lagrange multipliers. 
As illustrated in the example in Fig.1 of Ref.\cite{Pumplin1}, for
each $\lambda$ value, there is a singular minimum value of $\chi^2$ (Eq.(\ref{chi_lambda}))
as a function of  $F_2^{p(d)}$. 

In order to apply the Lagrange multiplier method to our SOM approach
we proceed as follows: 

\noindent {\it i)} we start from $F_2^{p(d)}$  as
determined by the SOMPDF code along with their  $\chi^2$ values.
These are calculated comparing all values of the $F_2^{p(d)}$, used in the SOMPDF
code and the corresponding values from experimental data;  

\noindent {\it ii)} we define the 
interval $\lambda \in [-200,200]$, and we increase $\lambda$
in increments of $10$;
we then generate sets of 
``pseudo experimental data" by shifting $F_2^{exp}$ for given $x$ and $Q^2$
values by $\pm \Delta F$, and we repeat the SOMPDF fit for each new data set.
The new structure functions are defined by a corresponding set 
of new individual PDFs, $F_2^{exp, \, NEW}$; 

\noindent {\it iii)} the difference between the individual 
PDFs from the limiting upper and lower $F_2^{exp, \, NEW}$ values define
then the Lagrange error for each of the individual PDFs for the original $F_2^{p(d)}$.

\begin{figure}
%\includegraphics[width=7.cm]{nnpdf_example.pdf}
%\hspace{-4.5cm}
%\includegraphics[width=9.cm]{pdfenvelope.pdf}
%\includegraphics[width=8.cm]{lagnew2.pdf}
\includegraphics[width=9.cm]{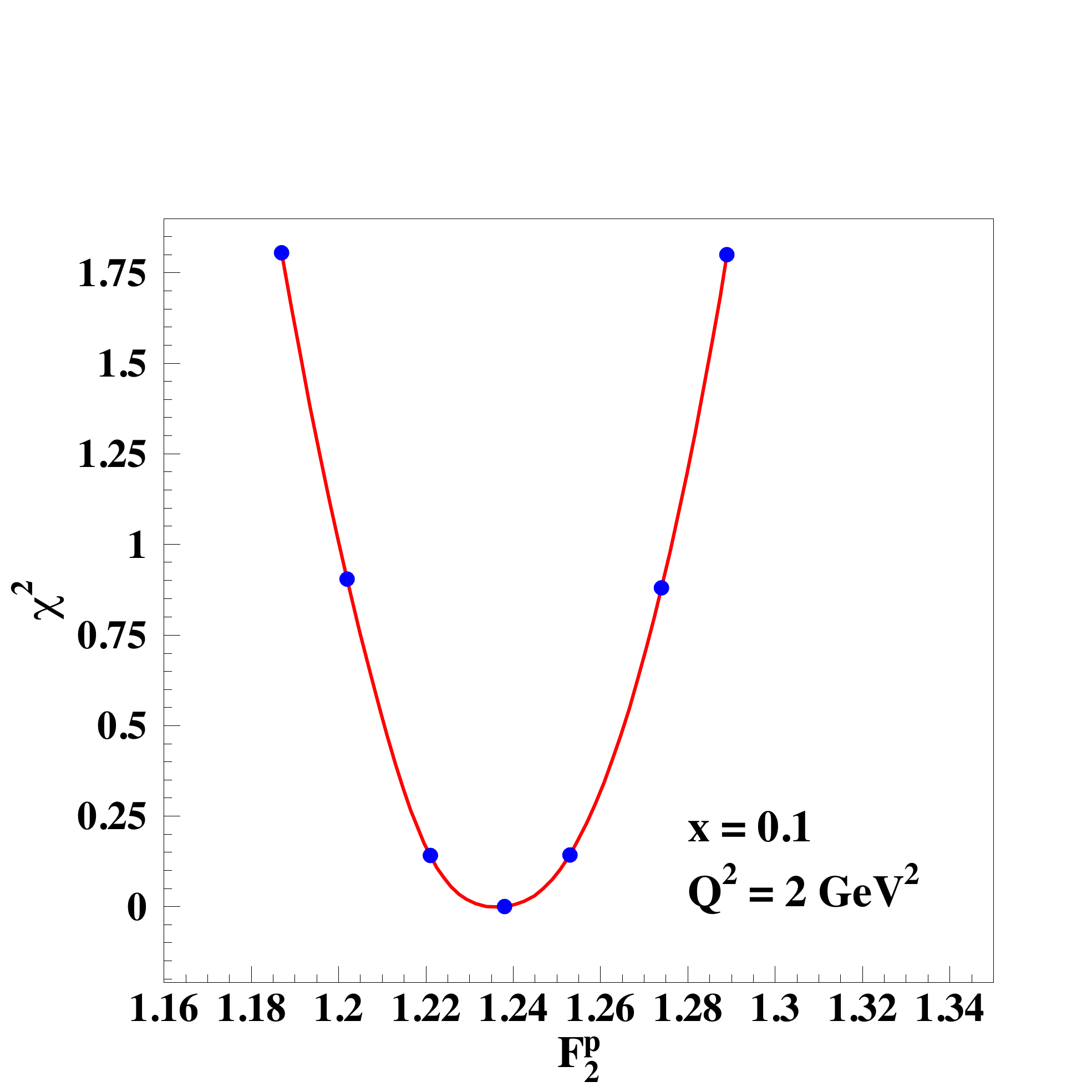}
\caption{(Color online) Illustration of the behavior of the minimum $\chi^2$ for the observable, $F_2^p$, obtained with the Lagrange multiplier's method. The dots correspond to $\lambda= 0, \pm 25, \pm 50$. The minimum $\chi^2$ was evaluated here in one particular kinematical bin: $x=0.1$, $Q^2=2$ GeV$^2$, corresponding to  just one of the terms in the sum in Eq.(\ref{chi2}). Similar graphs are obtained for all the other bins. Our analysis includes the experimental systematic error correlations from the data's normalization shown in Eq.(\ref{chi2}).}
\label{fig:lagrange}
\end{figure}

%Then I have the
%Chi Squared values I can plot as a function of the SOMPDF F2P/D
%values.  
The curves follow a 
parabolic shape.
%like there would be, for example, a singular minimum value for f(x) =$x^2$ as a function of x.  
The minima were calculated for all the $\lambda$ values,  and the differences between these minimum values
were used to determine the Lagrange error in 
$F_2^{p(d)}$.
%denoted here by $\Delta F$.
We show the minimum $\chi^2$ obtained for our choice of interval and $\lambda$ values plotted as a function of $F_2^{p}$ in Fig.\ref{fig:lagrange}.
%%%% FIGURES LAGRANGE
%%%%%
%%%%%

\begin{figure}
%\includegraphics[width=7.cm]{nnpdf_example.pdf}
%\hspace{-4.5cm}
%\includegraphics[width=9.cm]{pdfenvelope.pdf}
%\includegraphics[width=8.cm]{lagnew2.pdf}
\includegraphics[width=9.cm]{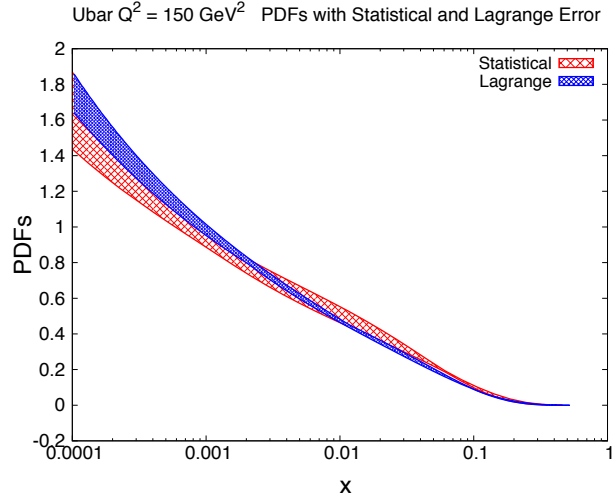}
\caption{(Color online) Illustration of the difference  between the the PDF uncertainty calculated with the Lagrange multipliers method and the statistical error analysis. Analogous results are obtained for other PDFs}
\label{fig:lagrange2}
\end{figure}

In Figure \ref{fig:lagrange2} we compare our results obtained with the Lagrange multipliers method with a simple statistical analysis applied to the conjoint set of final envelope PDFs and map PDFs. The PDF $\bar{u}$ is shown in the figure, analogous results are obtained for other PDFs. This type of statistical treatment resembles the early version of the NNPDF error analysis \cite{DelDebbio.2007}. However, instead of generating replicas of data, we just make use of the randomly generated set of PDFs (see Fig.\ref{fig:envelope}), and calculate the $\chi^2$ according to Eq.(\ref{chi2}). From Fig.\ref{fig:lagrange2} one can see that the uncertainty generated through the simple statistical analysis is different (it is larger) than the one obtained with the Lagrange multipliers method.  
%%%%%% FIGURE LAGRANGE VS. STATISTICAL

%%%%%%%%%%%%%%%%%%
%%%%%%%%%%%%%%%%%% RESULTS
%%%%%%%%%%%%%%%%%%
\section{Results}
\label{sec4}
We present our results by following the different stages of the SOMPDF analysis after the initialization part including the construction of the envelope in Section \ref{sec3A}. The map training and its clustering properties is shown in Section \ref{sec:clus}; an illustration of the SOM as a minimization procedure is given in Section \ref{sec:min}. Finally, the set of SOMPDFs with errors is given  in Section \ref{sec:pdfs}. 

%%%%% CLUSTERING MAPS
\subsection{Training and Clustering Properties}
\label{sec:clus}
In our approach the construction of the PDFs envelope is a fundamental component of the SOMPDFs initialization procedure (see Sections \ref{sec2C} and \ref{sec3A}). Once the maps are initialized, the training procedure begins. At the end of a training section the various PDFs are organized on the map. 

\begin{figure}
\includegraphics[width=9.cm]{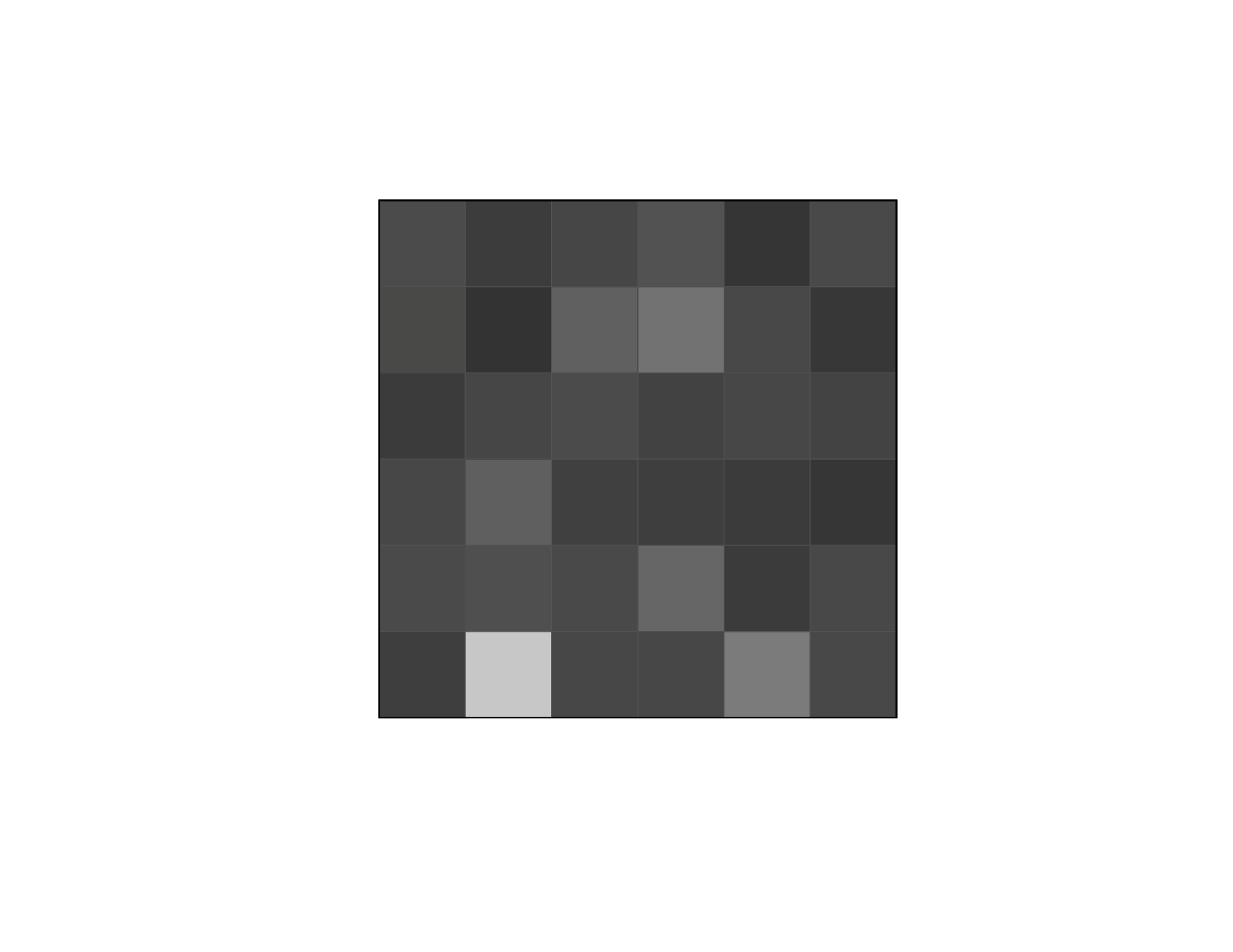}
\hspace{-1.2cm}
\includegraphics[width=9.cm]{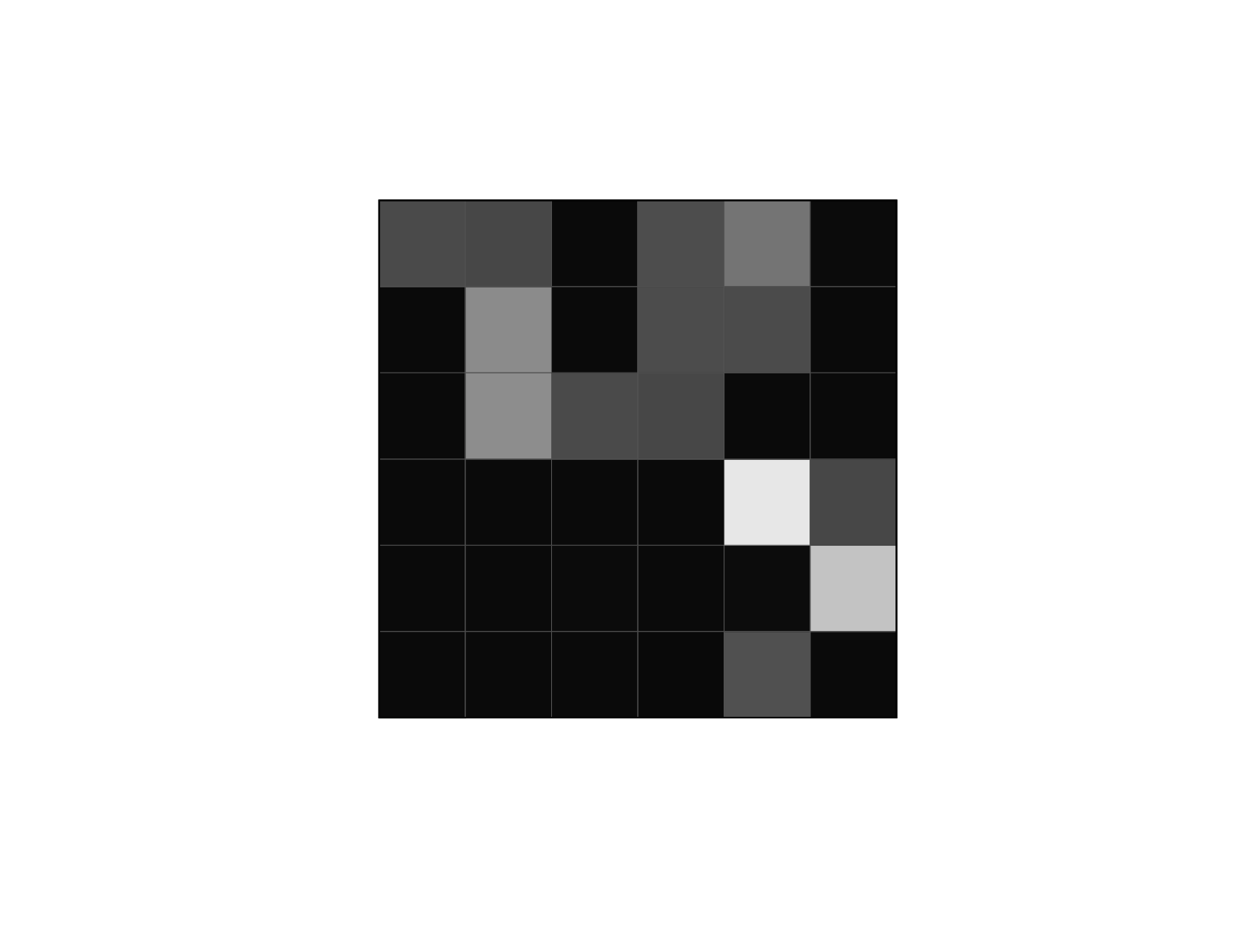}
\caption{Values of $\chi^2$ represented on a $6 \times 6$ map for the initial  (LHS) and final (RHS)  GA iterations. The values of $\chi^2$ are lowest for the darker squares. The clustering property of the map according to the $\chi^2$ is clearly visible.}
\label{chimap1}
\end{figure}

\noindent The clustering properties of the map are shown in Figure \ref{chimap1}. % \ref{chimap2}. 
In Fig. \ref{chimap1} each map cell is associated with a global $\chi^2$ value. The map on the LHS shows the initial step of the GA, while the one on the RHS shows the final step.
One can see that  while in the initial map the PDFs clearly do not group separately based on the $\chi^2$ criterion (they are homogenously distributed), in the final one the PDFs with lowest $\chi^2$ tend to cluster in the lower left corner. By analyzing the content of the cells in the lower left corner one can study the  various significant features of the PDFs that caused them  to group topologically on the map. 
%%%% FIGURE MAPs
%
%\begin{figure}
%\includegraphics[width=7.cm]{66chimap.pdf}
%\hspace{-4.5cm}
%\includegraphics[width=7.cm]{66chimap.pdf}
%\caption{Same as Fig.\ref{chimap1}, for the final GA iteration.}
%\label{chimap2}
%\end{figure}
%%%%%%%%%%%%%%%%%%%%%%%%%%%%%%%%%%%%%%%%%%%%%%%%%%%%%%%%%%%%

While the scope of the present paper is mainly to illustrate our method as a minimization procedure with a fully quantitative error analysis, future work will be directed at disentangling, through the use of SOMs different components of the DIS and deeply virtual exclusive and semi-inclusive processes. 

%{\bf here we need to show lego plots with no color side by side with flat color map of: 1) $\chi^2$, 2) $\bar{u}$ $\bar{d}$ difference, 3) $F_2^n/F_2^p$, 4) effects of cuts on data ($Q^2$, $x$).}

We conclude that by using the SOMPDF method we can discriminate more efficiently among different models, and among different features of models. More specific work in the large $x$ sector, where many effects such as target mass corrections, large $x$ resummation, higher twists, and nuclear dynamics corrections are simultaneously present, is in preparation \cite{Aska_prep}.

%%%% SUBSECTION: MINIMIZATION 
\subsection{Minimization}
\label{sec:min}
\begin{figure}
\includegraphics[width=9.cm]{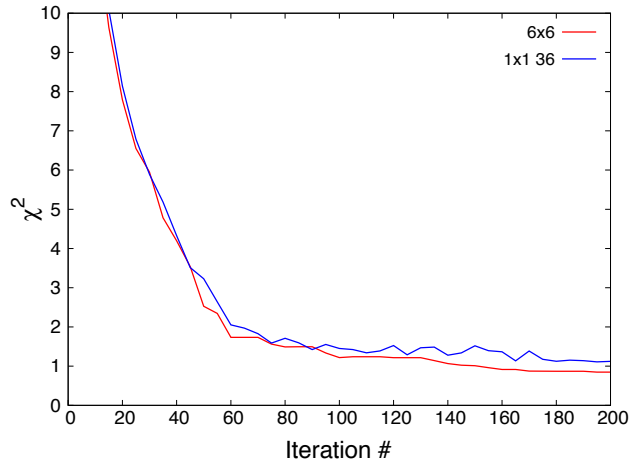}
%\hspace{-4.5cm}
\caption{(Color online) $\chi^2$ plotted vs. the  number of iterations for the GA. The two curves were obtained using a $6 \times 6$ maps and the GA only ($1 \times 1$ map).}
\label{chi1}
\end{figure}

Once the PDFs are represented on the map, the $\chi^2$ for each map cell is calculated according to Eq.(\ref{chi2}). Subsequent maps are run using the GA described in Section \ref{sec2C}.
As Fig.\ref{chi1} shows, at each GA iteration the $\chi^2$ decreases. We use the flattening of the $\chi^2$ as  a stopping criterium for our procedure.

Although it is clear that SOMs provide a unique tool for detecting uncertain features of the data, we also remark that the SOM's clustering properties are an important component specifically in the minimization procedure.  The question of  whether the GA would be  sufficient on its own, or of whether it would be playing a dominant role, independently from the map architecture is in fact ruled out in our approach. This would be equivalent, in the NNPDF procedure,  to using only their Monte Carlo sampling, while eliminating the neural network as an interpolator to obtain the continuous PDFs. The whole ensemble of non linear correlations among the functions would be completely missed. In the minimization procedure this would generate solutions for local minima.    
Nevertheless, we performed a quantitative check by  comparing results obtained using  $n \times n$ maps ($n>1$) with runs using only the GA, {\it i.e.} a $1 \times 1$ map with $n^2$ matchings. From Fig.\ref{chi1} one can see that the $\chi^2$ results are improved (even if only slightly) in the SOMPDF+GA treatment with respect to using GA only.  Furthermore, the final $n \times n$ map (see the $n=6$ case in Fig.\ref{chimap1} shows clustering, thus implying the existence of non-trivial non-linear error correlations. These are the errors which are taken into account in our analysis
(Section \ref{sec:err}). The other method give a $chi^2$ that comes close to a local minimum, but since each $1\times 1$ map is started from a new ensemble at each one of the $n^2$ iterations, error correlations are in this case disregarded.   

%%% CHI SQUARED FIGURE
%%%%%%%%%
%%%%%%%%%%
%%%%%%% SUBSECTION: ALL PDFs
%%%%%%%%
\subsection{PDFs with uncertainties}
\label{sec:pdfs}
\begin{figure}
\includegraphics[width=8.5cm]{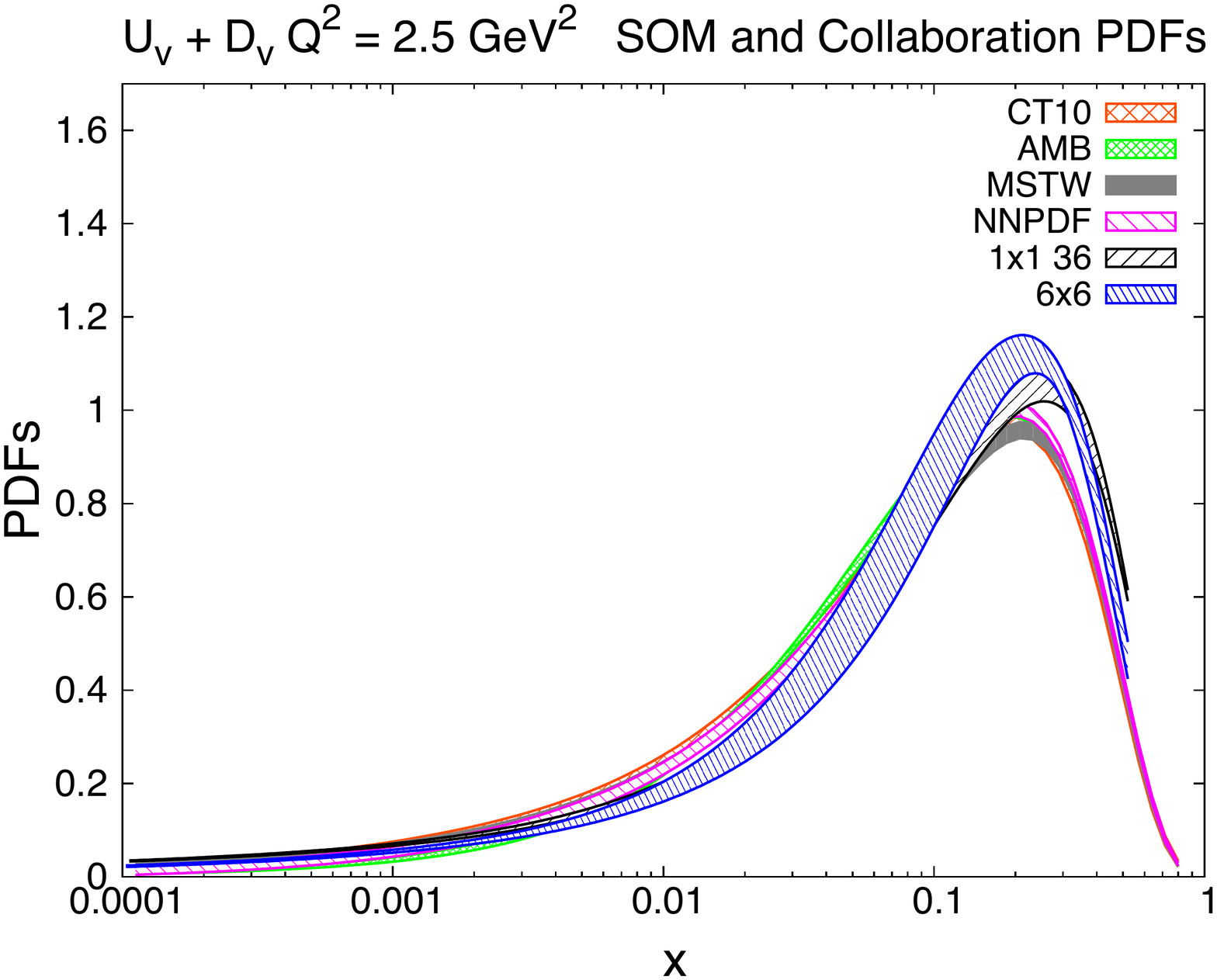}
\includegraphics[width=8.5cm]{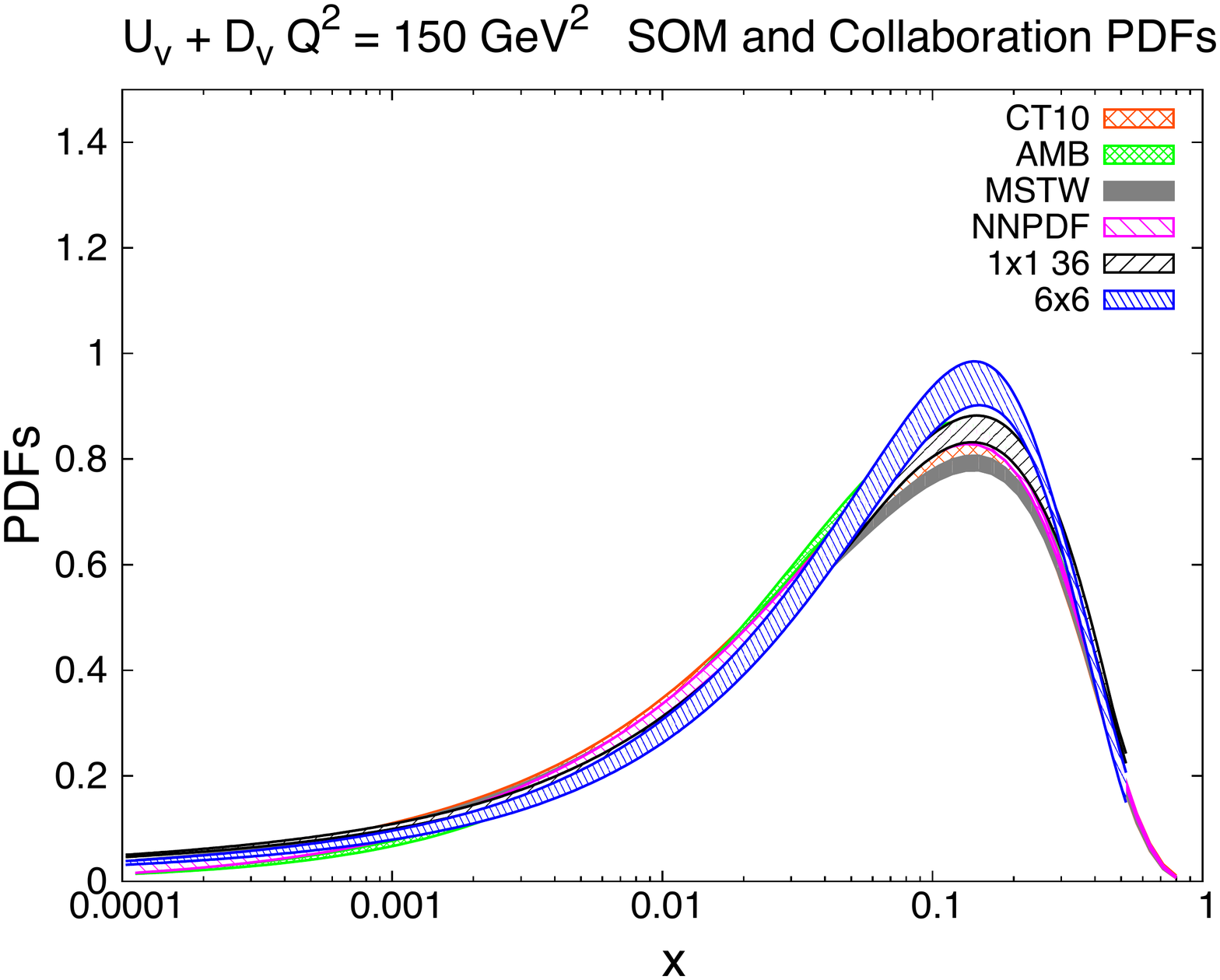}
\caption{Color online) NLO $u_v+d_v$ at $Q^2=2.5$ GeV$^2$ (LHS) and $Q^2=150$ GeV$^2$ (RHS).
SOMPDF1 results  were calculated with a $7 \times 7$ map, using 200 GA iterations. The uncertainty was calculated using the Lagrange multipliers method described in Section \ref{sec:err}. Shown for comparison are NLO results from \cite{CT10,Ball.2010,MSTW,ABM}.}
\label{fig:uvdv}
\end{figure}

\begin{figure}
\includegraphics[width=8.cm]{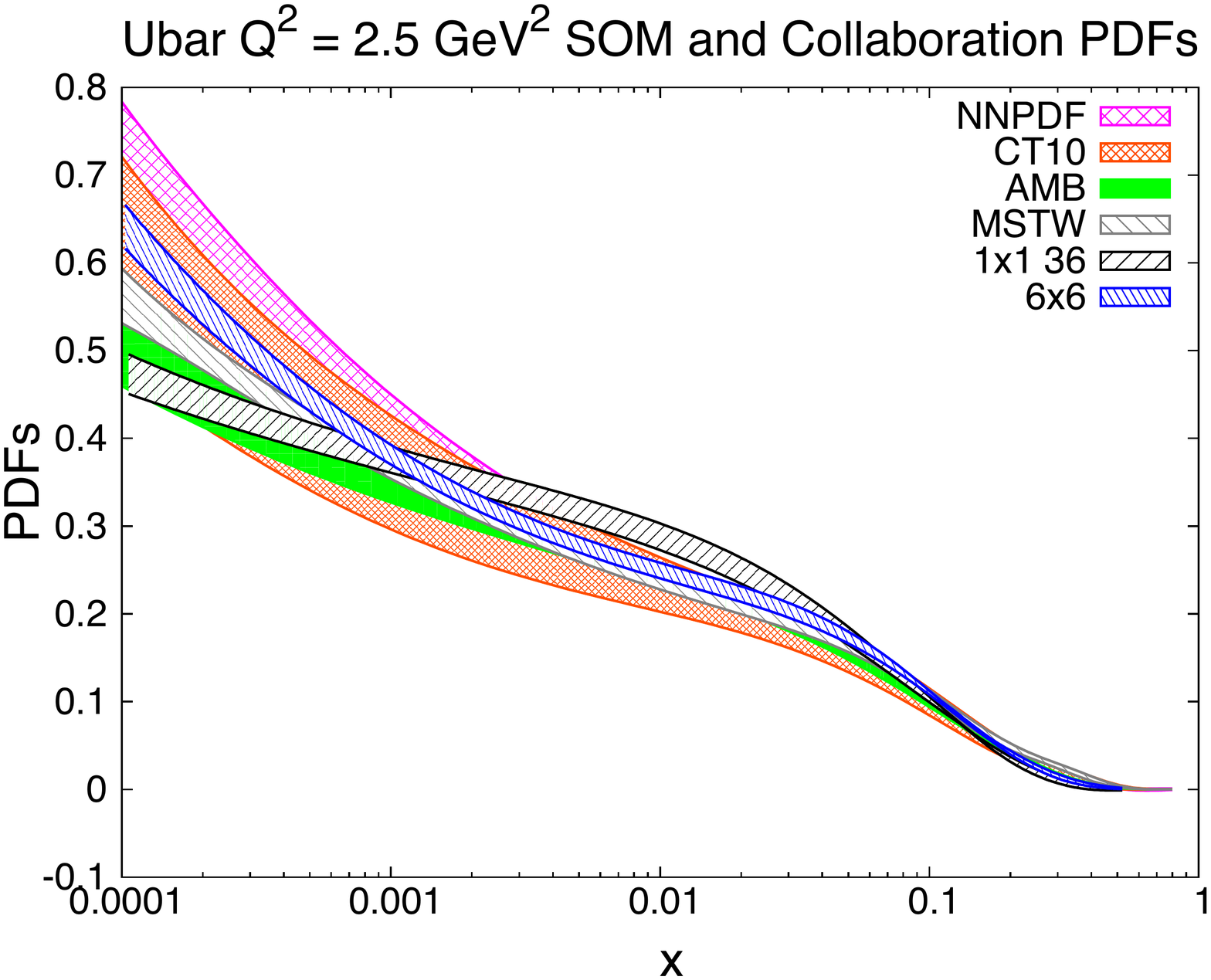}
\includegraphics[width=8.cm]{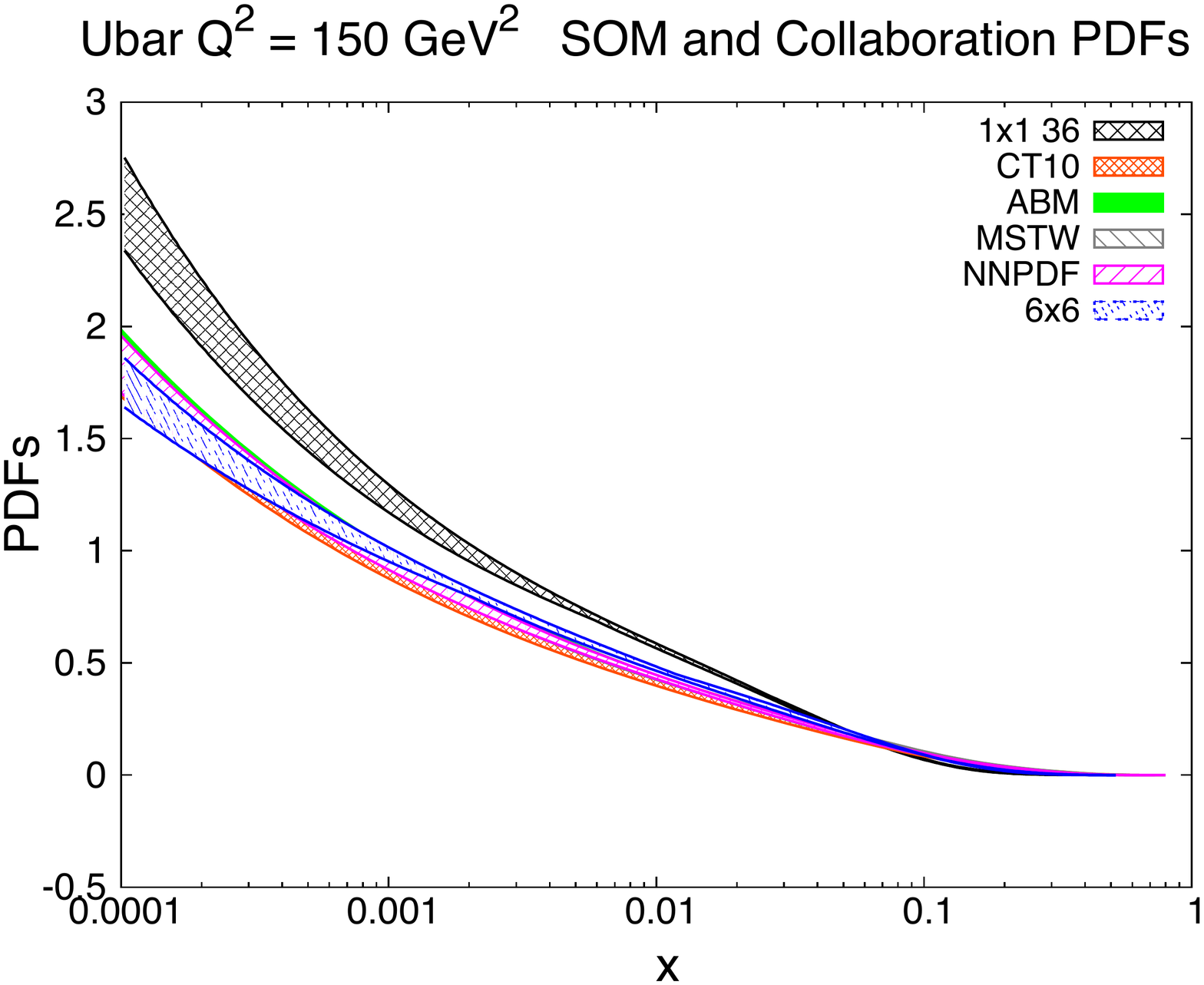}
\caption{(Color online) Same as Fig.\ref{fig:uvdv} for the $\bar{u}$  component.} 
\label{fig:ubar}
\end{figure}

\begin{figure}
\includegraphics[width=8.cm]{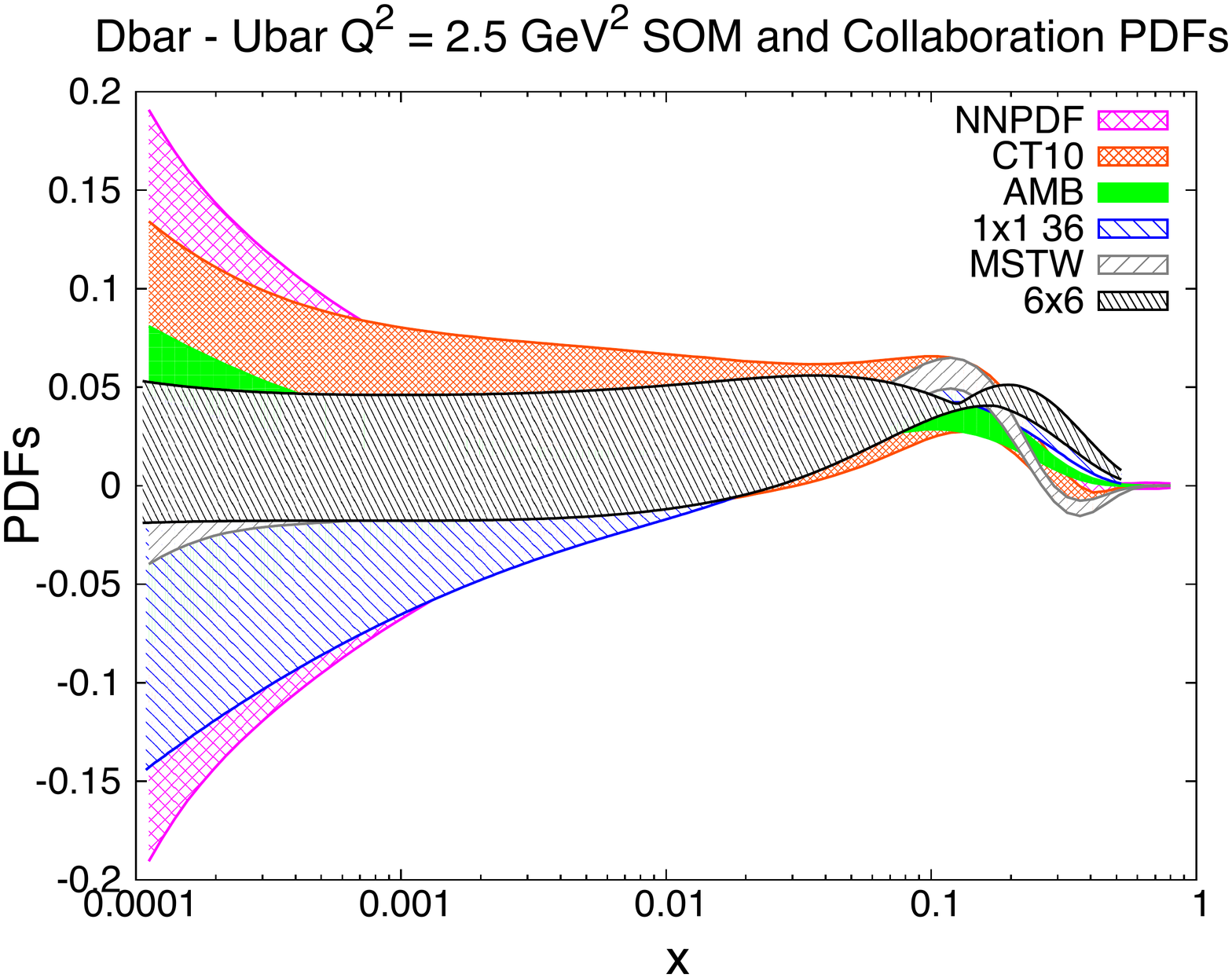}
\includegraphics[width=8.cm]{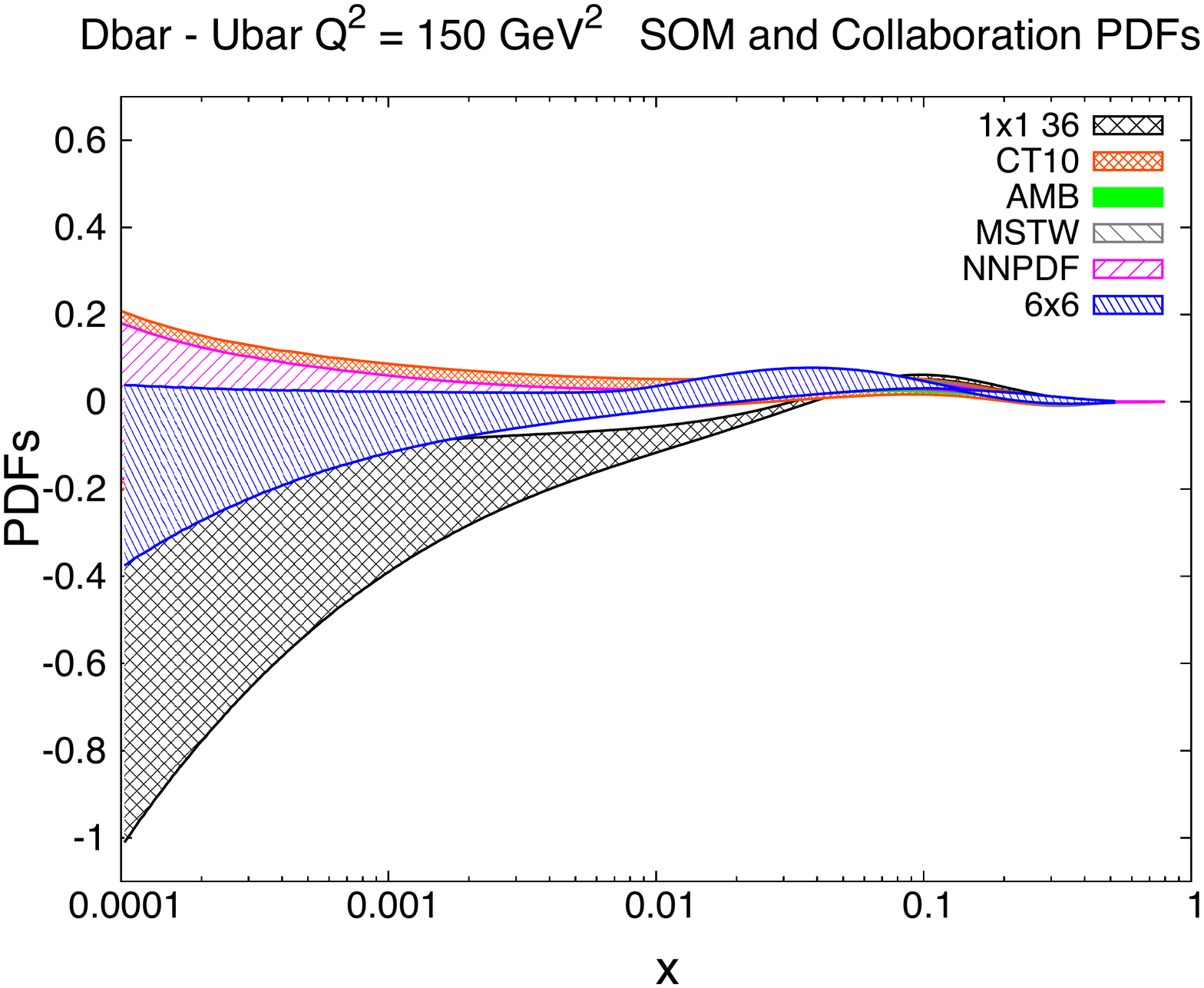}
\caption{(Color online) Same as Fig.\ref{fig:uvdv} for $\bar{d}-\bar{u}$.} 
\label{fig:ubdb}
\end{figure}

\begin{figure}
\includegraphics[width=8.cm]{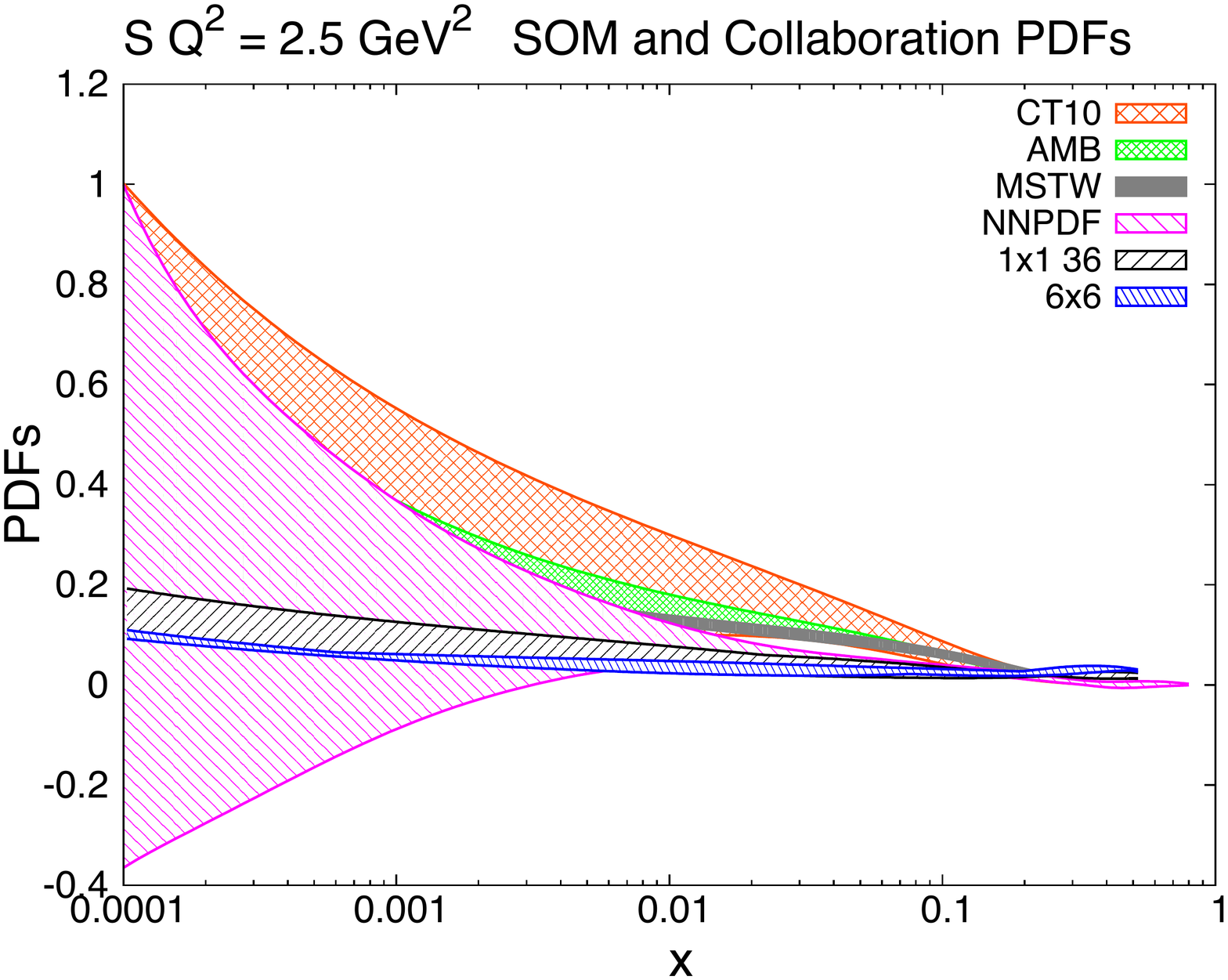}
\includegraphics[width=8.cm]{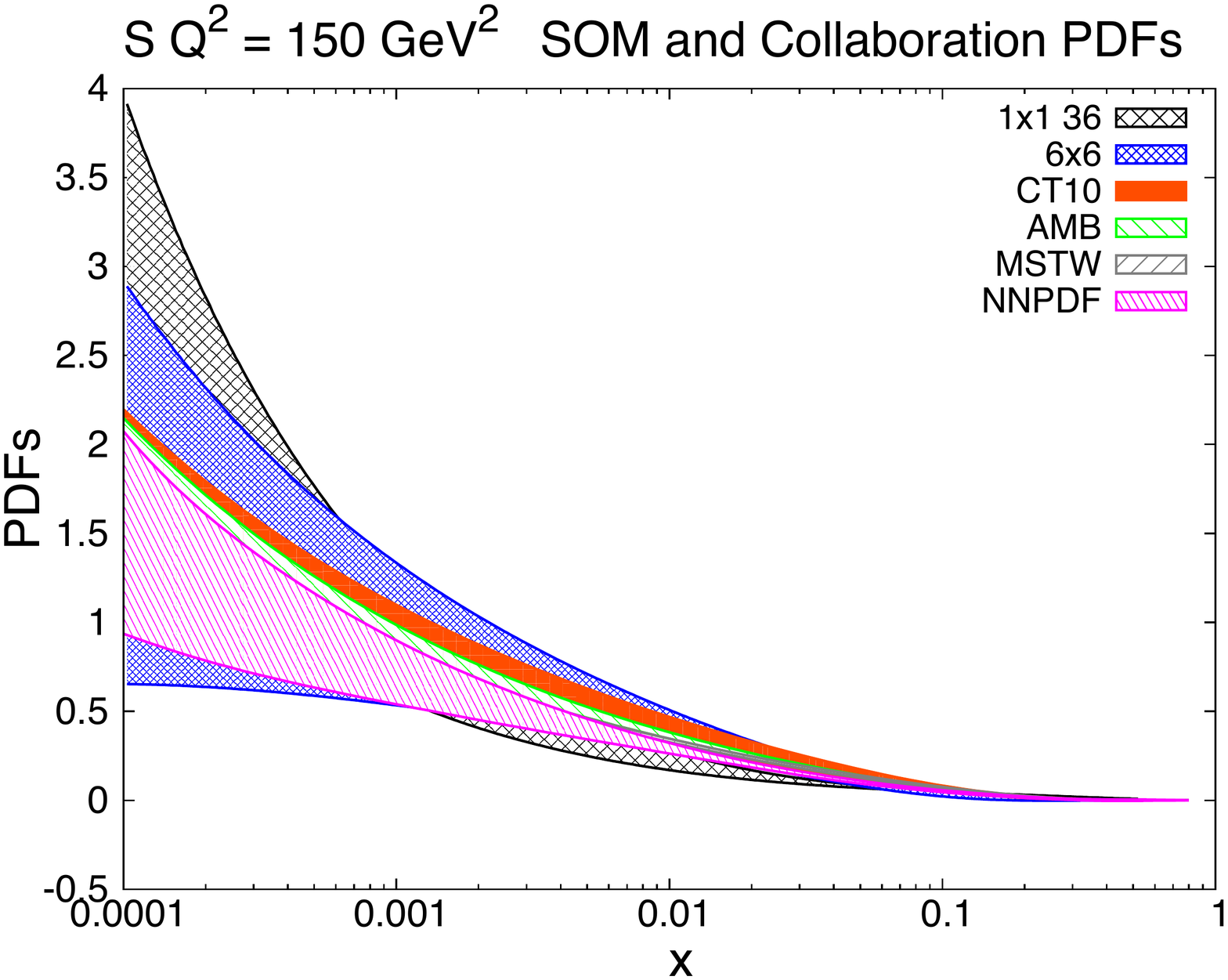}
\caption{(Color online) Same as Fig.\ref{fig:uvdv} for the $s$ component.}
\label{fig:s}
\end{figure}

\begin{figure}
\includegraphics[width=8.cm]{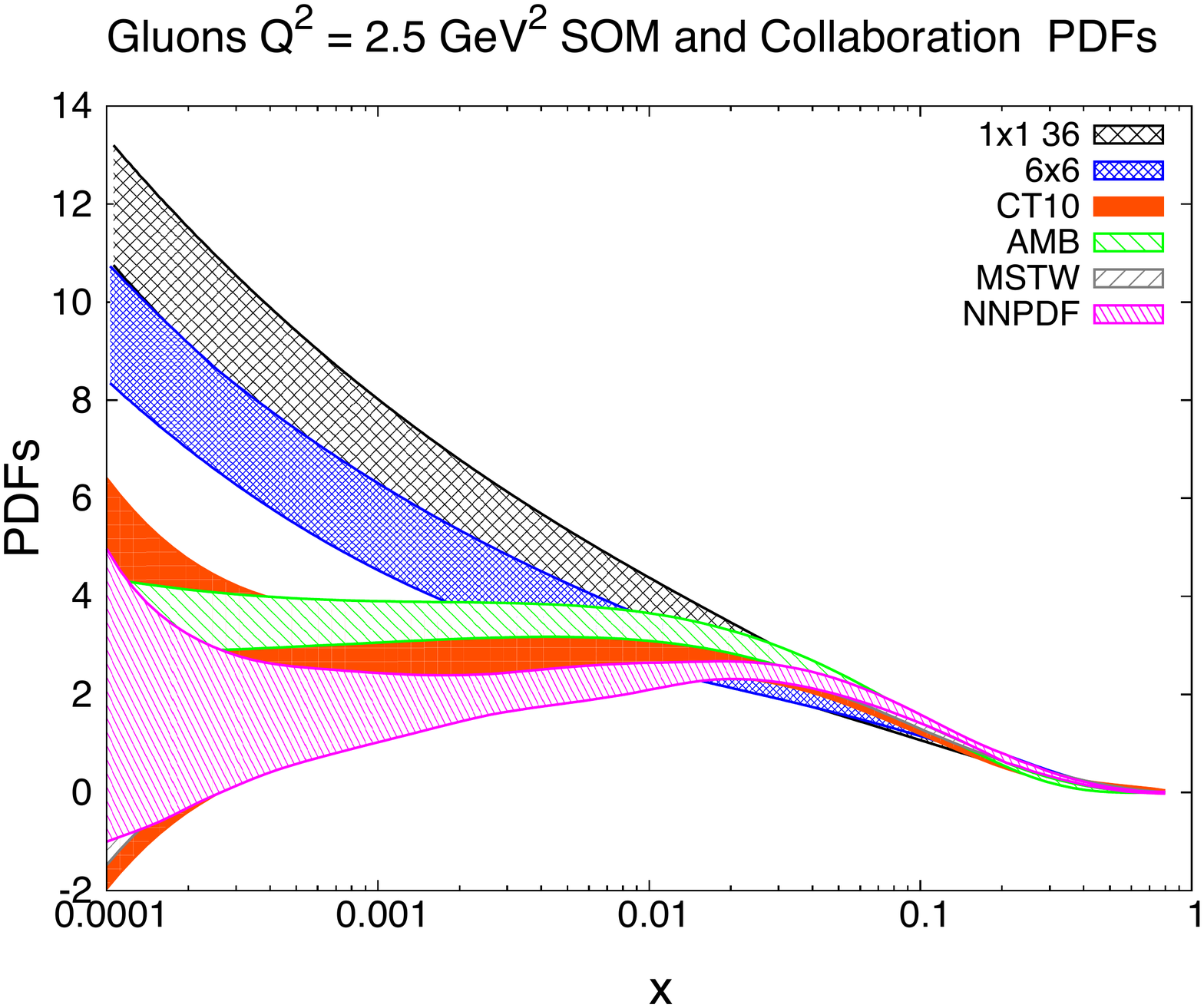}
\includegraphics[width=8.cm]{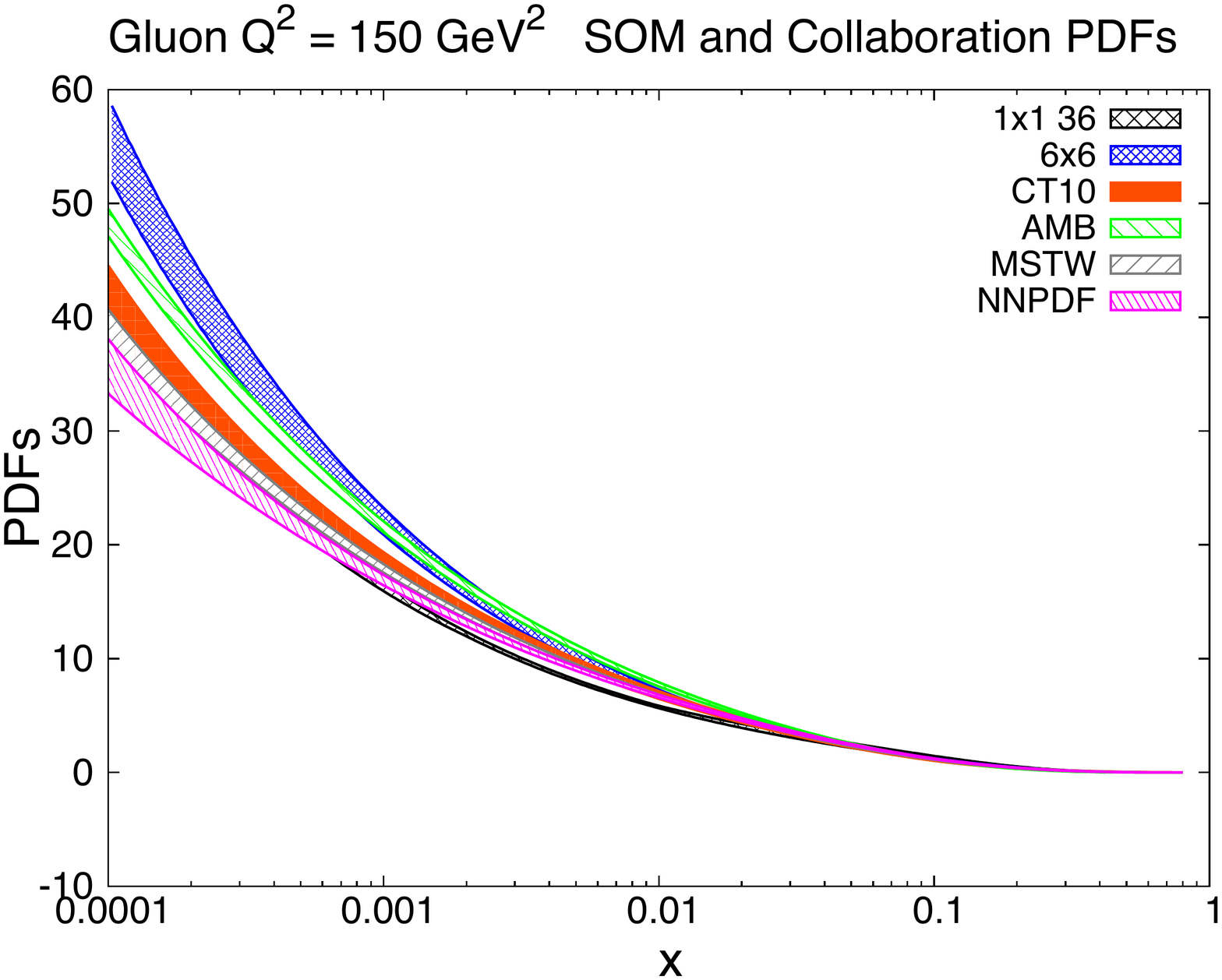}
\caption{(Color online) Same as Fig.\ref{fig:uvdv} for gluons.}
\label{fig:gluon}
\end{figure}

We now present results for the SOMPDF1 parton distributions. The separate PDFs are shown in Figures  \ref{fig:uvdv}, \ref{fig:ubar}, \ref{fig:ubdb}, \ref{fig:s}, \ref{fig:gluon}. In each figure we compare  our analysis to other available NLO parameterizations \cite{CT10,Ball.2010,MSTW,ABM}. We show results obtained with a $6 \times 6$ map, with 200 GA iterations. The error bands were calculated using  the Lagrange multipliers method (see Section \ref{sec:err}).  Results obtained using the GA only, skipping the map construction, are also shown in the figures (labeled as $1\times1$). The GA alone  seemingly succeeds in minimizing the $\chi^2$ with an efficiency only slightly lower than  the full map algorithm. It however misses the important nonlinear statistical correlations among high-dimensional data which are accounted for in the self-organization process.
\footnote{This would be comparable to omitting the interpolating properties in NNPDF \cite{Ball.2010}.} 
Therefore, the GA selection by itself is likely to miss the true solution, and to reach instead  a local minimum. This can be seen, for instance, by inspecting in Fig.\ref{fig:ubar} (right). 

%%%%% PDFS with errors FIGURES
%%% UV DV

%%%% Ubar 

%%%% Dbar Ubar 

%%%% S

%%%% gluons

\begin{figure}
\includegraphics[width=12.cm]{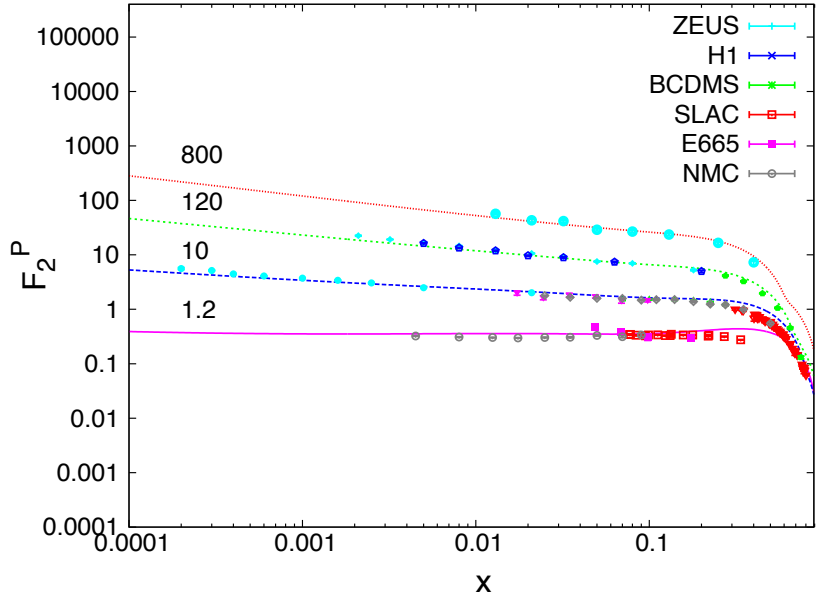}
\caption{(Color online) Structure function $F_2^p$ plotted vs. $x$ at different $Q^2$ values ($Q^2=1.2, 10, 120, 800$ GeV$^2$). Experimental data from Refs.\cite{SLAC,BCDMS,NMC,H1,ZEUS}.}
\label{fig:F2P}
\end{figure}

In Figure \ref{fig:F2P} we show a comparison of our results with experimental data \cite{SLAC,BCDMS,NMC,H1,ZEUS}. The errors were calculated using the Lagrange multipliers method. The usage of SOMs allows us to account for non linear correlations in the uncertainties.  
%%%%%%% FIGURE
%%%%% F2
%%%%%%%%%%%%%%%

\section{Conclusions and Outlook}
\label{sec5}
We presented a new parametrization of  PDFs from deep inelastic scattering data based on a new type of neural networks, the SOMs. Similarly to NNPDF  we avoid the bias that is associated with the choice of a specific parametric functional form for the PDFs which characterizes standard/non-ANN-based, global analyses.
The way a generic ANN-based ``unbiased" approach functions, however, is by  increasing the flexibility of the functional form used to fit the data. A much larger number of parameters is introduced (given in the NNPDF case by the weights). The price to pay for the extreme flexibility and unbiasedness is given by the very large error bars -- essentially an almost complete  indetermination -- which are obtained for certain observables and kinematical regions where the data are scarce. Through the self-organizing procedure which characterizes the SOMPDF fit, we restore, instead, some of the predict power compared to generic ANN-based fits by exploiting latent correlations existing among data, while retaining the unbiasedness feature of the neural networks. The underlying principle we exploit is the unsupervised learning aspect  of the SOM.

In this paper we showed that our method works quantitatively. In order to accomplish this, some major improvements were applied to our initial exploratory analysis in Ref.\cite{Carnahan} where the idea of SOMPDF was first introduced.  We now provide a parameterization for PDFs at NLO, SOMPDF1, with an uncertainty analysis which was carried out using the Lagrange multipliers method. Similarly to how several ``conventional" PDFs fitting procedures, differing among themselves in several aspects, have been providing guidance for data analyses, our method can parallel and support the ANN based effort. 

%%% Outlook
Applications of our analysis are in progress, which address specific aspects of PDFs such as their large $x$ behavior,  
or nuclear effects that are  important for the interpretation of neutrino experiments, and  that are well known to affect  non trivially the extraction from the data of the proton and neutron PDFs. 
Our method will be, however, most fruitfully applied to
semi-inclusive and exclusive hard scattering experiments 
that  are sensitive to additional degrees of freedom of hadronic structure, and that are characterized, therefore, by a higher degree of complexity. 

For instance, experiments performed at DESY, Jefferson Lab, and CERN, have been and currently are designed to disentangle eight 
leading twist Transverse Momentum Distributions (TMDs), which depend on the additional variable of partonic intrinsic transverse momentum, and eight Generalized Parton Distributions (GPDs) which are related to  transverse spatial partonic configurations.An even larger number of higher twist distributions play an important role in the interpretation of experimental data. 
Furthermore, to extract  these functions from data is challenging since they have to be disentangled from integrated quantities. This aspect, added to the increased number of  degrees of freedom per function, makes it a daunting task to perform fully quantitative fits including a precise treatment of the error.

SOMPDFs provide an ideal tool for analyzing these rather elusive observables owing to their various features which were discussed in this paper. They allow one to extrapolate to regions where data are  scarce due to the unsupervised learning algorithm. They provide an advantage in the visualization of data features. Finally,  they provide a way to obtain enhanced information on complex, multivariable dependent phenomena. 

\acknowledgements
We thank Dan Perry (Perahya) for his involvement in the initial stage of the analysis, Swadhin Taneja for comments on the Lagrange multiplier's analysis, and Donal Day for 
support and discussions about data sets. We are also indebted to UVACSE for assistance with code development and resource utilization. 
This work was supported by the U.S. Department
of Energy grants DE-FG02-01ER4120 (E.M.A. and  S.L.), and DE-FG02-96ER40950 (E.M.A.).
%%%%%%%%%%%%%%%%%%%%%%%%%%%%%%%%%%%%%%%%%%%%%%%%%%%
%%%%%%%%%%%%%%%%%%%%%%%%%%%%%%%%%%%%%%%%%%%%%%%%%%%
%%%%%%%%%%%%%%%%%%%%%%%%%%%%%%%%%%%%%%%%%%%%%%%%%%%

\end{document}